\documentclass[preprint,authoryear,12pt]{elsarticle}

\usepackage{graphics}
\usepackage{graphicx}
\usepackage{epsfig}
\usepackage{amssymb}
\usepackage{amsthm}
\usepackage{bm}
\usepackage{amsmath}
\usepackage{color}

\biboptions{}

\journal{Journal of the Mechanics and Physics of Solids}
\begin{document}

\begin{frontmatter}



\title{Modified Multiplicative Decomposition Model for Tissue Growth: Beyond the Initial Stress-free State}


\author[1]{Yangkun Du} 
\author[2,3,4]{Chaofeng L\"{u}\corref{cor1}} 
\ead{lucf@zju.edu.cn}
\author[1,3,4]{Weiqiu Chen}
\author[1,5]{Michel Destrade}

\cortext[cor1]{Corresponding author}
\address[1]{Department of Engineering Mechanics, \\ Zhejiang University, Hangzhou 310027, P. R. China;\\[6pt]}
\address[2]{Department of Civil Engineering,\\ Zhejiang University,  Hangzhou 310058, P.R. China; \\[6pt]}
\address[3]{Key Lab of Soft Machines and Smart Devices of Zhejiang Province, \\ Zhejiang University, Hangzhou 310027, P.R. China; \\[6pt]}
\address[4]{Soft Matter Research Center, \\ Zhejiang University, Hangzhou 310027, P. R. China; \\[6pt]}
\address[5]{School of Mathematics, Statistics and Applied Mathematics,\\ NUI Galway, University Road, Galway, Ireland.}

\begin{abstract}
	
The multiplicative decomposition model is widely employed for predicting residual stresses and morphologies of biological tissues due to growth. However, it relies on the assumption that the tissue is initially in a stress-free state, which conflicts with the observations that any growth state of a biological tissue is under a significant level of residual stresses that helps to maintain its ideal mechanical conditions. Here, we propose a modified multiplicative decomposition model in which the initial state (or reference configuration) of a biological tissue is endowed with a residual stress instead of being stress-free. 

Releasing theoretically the initial residual stress, the initially stressed state is first transmitted into a virtual stress-free state, thus resulting in an initial elastic deformation. The initial virtual stress-free state subsequently grows to another counterpart with a growth deformation, and the latter is further integrated into its natural configuration of a real tissue with an excessive elastic deformation that ensures tissue compatibility. With this decomposition, the total deformation arising during growth may be expressed as the product of elastic deformation, growth deformation and initial elastic deformation, while the corresponding free energy density should depend on the initial residual stress and the total deformation. Three key issues including the explicit expression of the free energy density, the predetermination of the initial elastic deformation, and the initial residual stress are addressed. 

Finally, we consider a tubular organ as a representative example to demonstrate the effects of the proposed initial residual stress on stress distribution and on shape formation through an incremental stability analysis. Our results suggest that the initial residual stress exerts a major influence on the growth stress and the morphology of biological tissues. The model bridges the gap between any two growth states of a biological tissue that is endowed with a certain level of residual stresses.

\end{abstract}


\end{frontmatter}




\section{Introduction}


It has long been recognized that growth, death and all other bio-behaviors of living matter are controlled by a combination of genetic and epigenetic factors including biochemistry, bioelectricity and biomechanics \citep{RN260, RN575}. One of the most accepted biomechanical epigenetic factor in living matter may be its internal mechanical stress, which is also called residual stress in unloaded conditions \citep{RN478, RN570, RN576, RN572}. It exists in all real living matter such as ripe fruits, tree trunks, blood vessels or solid tumors, and is generally induced by non-uniform plastic deformation, surface modification, material phase changes and/or density changes \citep{RN488, RN281, RN572, RN503}.  Though these residual stresses are locally self-equilibrating in mechanics, they still serve some special functions, as they, for example, influence the morphogenesis, growth rate and internal mechanical condition of bio-tissues (e.g. \cite{RN90, RN523, RN98, RN451}). 

The first presentation on the relationship between stress and bio-behavior goes back to the German anatomist and surgeon Julius Wolff  \citeyearpar{ RN595} who showed that healthy bone creates structural adaptation where external loads are placed. Then \cite{ RN456} proposed the functional adaptation concept that stress should be regarded as a functional stimulus to growth and remodeling. Later, Fung and collaborators \citep{ RN282, RN523} used the opening angle method to quantify residual stress in arteries. 

\begin{figure}
	\centering
	\includegraphics[width=0.8\textwidth]{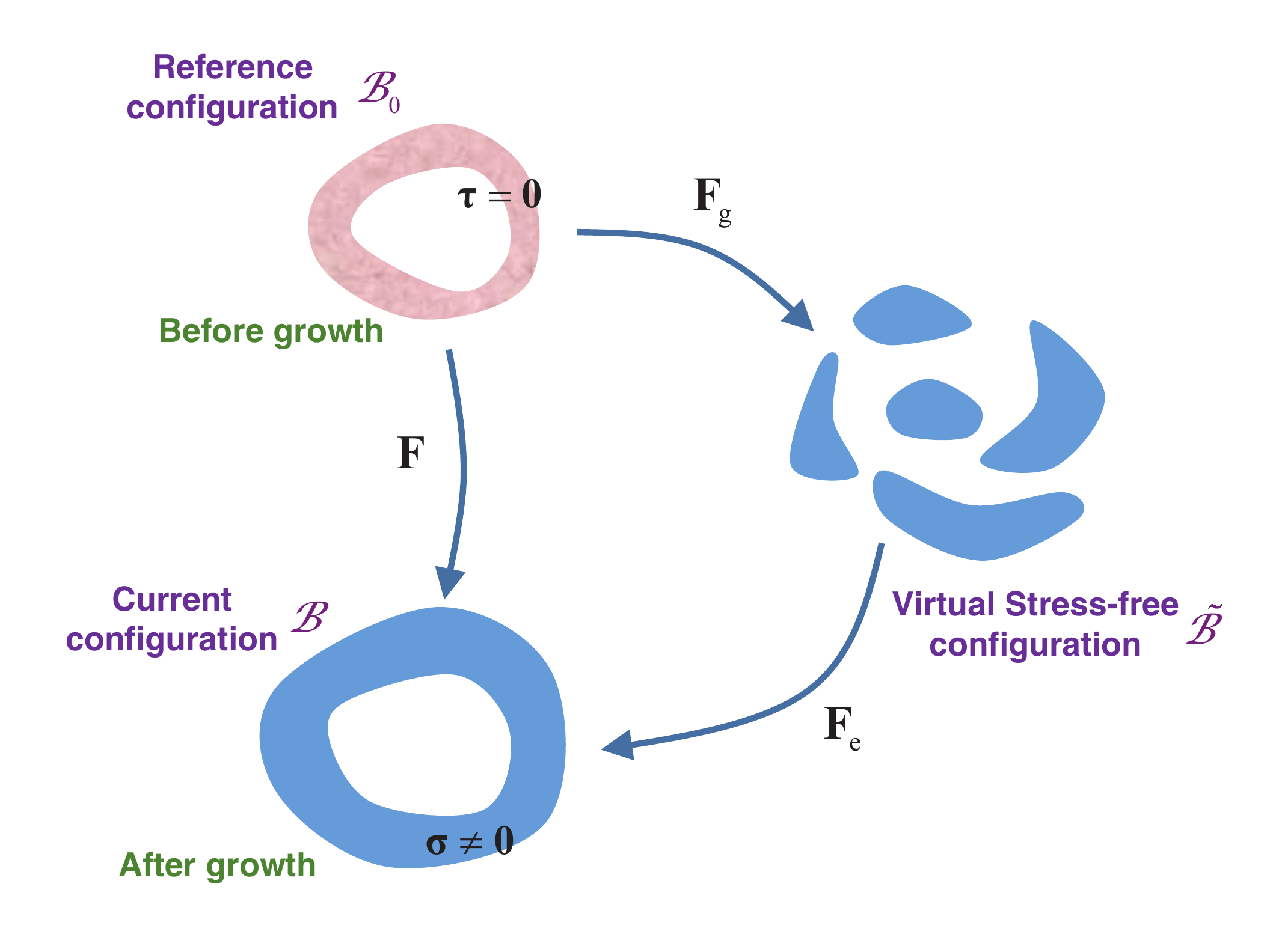}
	\caption{ The classical multiplicative decomposition proposed by Rodriguez}	
	\label{Fig:1}
\end{figure}

An explanation on the origin of residual stresses in living bio-tissues was first presented theoretically by \cite{RN101} via the multiplicative decomposition (MD) method. They showed that residual stresses in a bio-tissue are created by heterogeneous growth and can be calculated from a given growth gradient tensor. The constrained growth deformation is decomposed into unconstrained growth deformation and pure elastic deformation (Figure \ref{Fig:1}) with the relation $\mathbf{F}={{\mathbf{F}}_{{e}}}{{\mathbf{F}}_{{g}}}$, where $\mathbf{F}$ is the total deformation, ${{\mathbf{F}}_{{e}}}$ the pure elastic deformation, and ${{\mathbf{F}}_{{g}}}$ the growth deformation. From a modeling standpoint, this explanation is concise but powerful enough to predict the growth-induced residual stresses. The MD model was subsequently widely employed to solve many biomechanical problems related to growth process (e.g. \cite{RN529, RN612, RN98, RN374, RN503,wang2017}). 

\begin{figure} 
	\centering
	\includegraphics[width=1\textwidth]{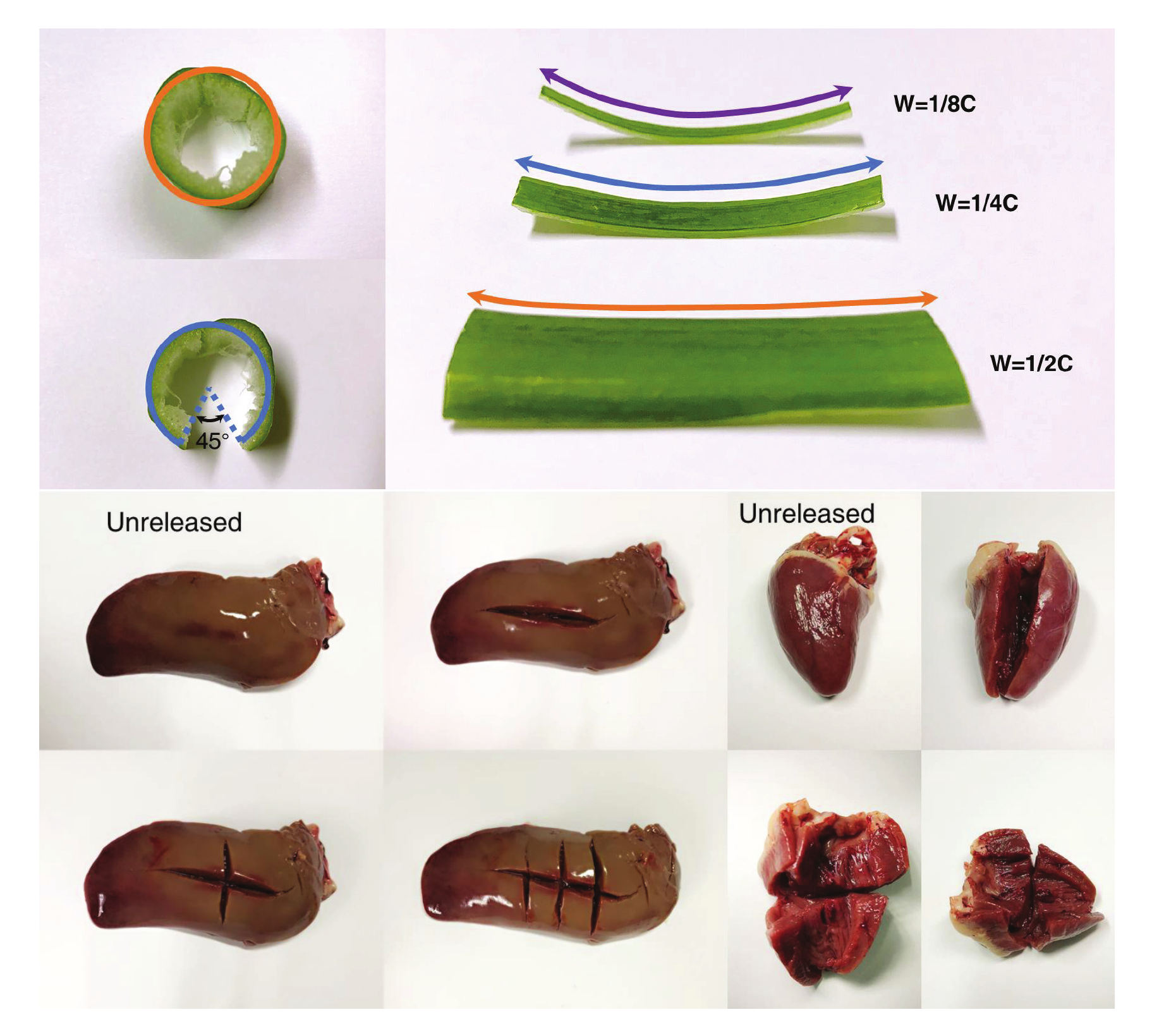}
	\caption{ Evidence of residual stress in biological tissues. (a) Demonstration by the opening angle of a scallion ring after cutting along the axial direction; (b) a different deformation induced by releasing residual stresses in scallion strips, where $W$ is the width of the strip and $C$ is the scallion circumference. Influence of the number and direction of cuts to release residual stress in (c) a duck liver and (d) a duck heart.}
	\label{Fig:2}
\end{figure}

One prerequisite of the MD model is that the reference configuration must be a \textit{stress-free state},  which ensures that the growth process is under the unconstrained condition. To study the growth process of bio-tissues starting from an arbitrary stage, the initial reference state should be properly defined. As evidenced by cutting experiments, many bio-tissues still exhibit large amounts of residual stresses even when the external loads are removed, as seen with a cut scallion, a duck heart or liver, see Figure \ref{Fig:2}, and also with cut arteries or weasands \citep{RN98}. 

An approximate method to access the stress-free state is by cutting the material to remove constraints from surrounding tissues, which is the basic idea behind the \textit{opening angle method} (e.g. \cite{RN282, RN389, RN572}). 
However, the extent to which residual stresses can be released depends significantly on the number and direction of the cuts (Figure \ref{Fig:2}b). 
In effect, an entirely stress-free state for a real living body can only be accomplished by an infinite number of cuts to release all residual stresses held by the neighboring regions. 
In practice, this ultimate discrete state is impossible to reach for real living matter \citep{ RN572}. This suggests that stress-free state may not be achievable for real bio-tissues, and, therefore, the stress-free reference configuration in theoretical modelling may not be appropriate. From this point of view, the conventional MD model \citep{RN101} is inadequate for predicting the growth of biological tissues starting from an arbitrary growth state that is regarded as the stress-free reference configuration.

A more practical reference configuration containing residual stresses is needed, a fact which has already been well recognized and achieved in mechanics of materials with initial residual stresses. For example, using the concept of mathematical limitation, Hoger and collaborators  \citep{ RN483, RN492} put forward the idea of a \textit{virtual stress-free state} to deduce the residual stress $\mathbf{T}$ in the form of $\mathbf{T=}\bar{\Im }\left( \mathbf{F,\bm{\tau}  } \right)$, where $\bar{\Im }$  is a mapping from the initially stressed  state to the current configuration, $\mathbf{F}$ is the elastic deformation gradient tensor and $\bm{\tau}$ is the initial stress. The virtual stress-free configuration is adopted only to give a physical interpretation for the mathematical derivation of the stress. Hence, the constitutive equation is directly established in the initially stressed state, and the reference configuration for a large deformation is no longer constrained to be the stress-free state. 

As an extension of the initial residual stress theory \citep{RN483, RN492},  \cite{ RN255} proposed a diagram to show the kinematic description for initially stressed growing matter, in which the growth process from an initially stressed state is decomposed into a sequence of releasing initial stresses, growing unconstrained in a  stress-free field, and finally yielding the residual stress. 
This is a more general description of growth, in which, however, initial stresses are created by an initial external load and can be entirely released by removing it. Hence Skalak et al.'s \citeyearpar{RN255} description of initially stressed growing matter does not consider existing initial residual stress. 

\color{black} \cite{goriely2007}  proposed a cumulative growth law to analyze the cumulative effect of residual stress during a large growth deformation. The total growth deformation was divided into many small steps that are further decomposed by the MD model. Except for the first step, any accumulative step grows from a residually stressed state, which provides a way to analyze the influence of residual stress on the growth process. However, these residual stresses can only be determined from the prior growth step, and, especially in the first step, it is still necessary to assume an initial stress-free state that is very difficult to prescribe in real biological tissues.\color{black}   

Later,   \cite{ RN452} proposed a free energy density $\psi {=}\psi \left( \mathbf{F},\bm{\tau} \right)$ as a function of ten tensor invariants to derive the residual stress created by the elastic deformation from an initially stressed state.  Then  \cite{ RN389} showed that the initial stress symmetry (ISS) condition can demonstrate the natural rationality of using the virtual stress-free state.

To describe the growth process of a bio-tissue from any state with residual stresses, we propose  an initially stressed reference configuration without external load from which the tissue grows to a current configuration. With this basic idea, we may avoid the drawbacks of using an unpractical virtual stress-free configuration as an initial state of growth in the conventional MD model \citep{RN101}. However, there remain two challenges for the current approach of modelling growth: one is how to construct a free energy density for the growth with initially stressed configuration, the other is how to perform a complete analysis for the growth process. These challenges will be addressed.

The paper is organized as follows. Section 2 outlines the detailed framework of the modified multiplicative decomposition (MMD) model, including the modified kinematic description, the virtual stress-free configurations, and the governing equations. 
Three key issues involved in the MMD model are also reviewed and primary resolutions are described. 
In Section 3, we consider a residually-stressed, growing neo-Hookean material as an illustrative example and derive the corresponding constitutive equations. 
Typically, we perform an inverse analysis to achieve the initial elastic deformation. Then, we analyze theoretically in Section 4 the growth-induced residual stresses for a tubular organ for which the distribution of initial residual stress is determined according to the self-equilibrium conditions. In Section 5, we analyze the growth-induced morphology of an initially stressed tubular organ using incremental theory to signal the onset of wrinkles. Both the influence of initial stress and initial wall thickness on the critical differential growth extent and the stability pattern are investigated. Finally, we discuss the significance of the work and draw some conclusions.


\section{Basic Theory}
\label{section2}



\subsection{Modified kinematic description of growth}


In the conventional MD growth model, the growth process is decomposed into two successive steps: first the mass accumulation process from the reference configuration ${{\mathcal{B}}_{{0}}}$ to the virtual stress-free configuration $\tilde{\mathcal{B}}$, second the elastic deformation process from $\tilde{\mathcal{B}}$ to the current configuration $\mathcal{B}$, modulating the morphologic compatibility and inducing residual stresses (Figure \ref{Fig:1}). 
As a result, the entire process of deformation can be expressed mathematically by $\mathbf{F}={{\mathbf{F}}_{{e}}}{{\mathbf{F}}_{{g}}}$. 
As mentioned previously, the reference configuration ${{\mathcal{B}}_{{0}}}$ in the conventional MD model is a stress-free state that can only be achieved by cutting the solid into an infinite number of discrete elements rather than be prescribed in a real bio-tissue as a continuous configuration. 

\begin{figure}
	\centering
	\includegraphics[width=.9\textwidth]{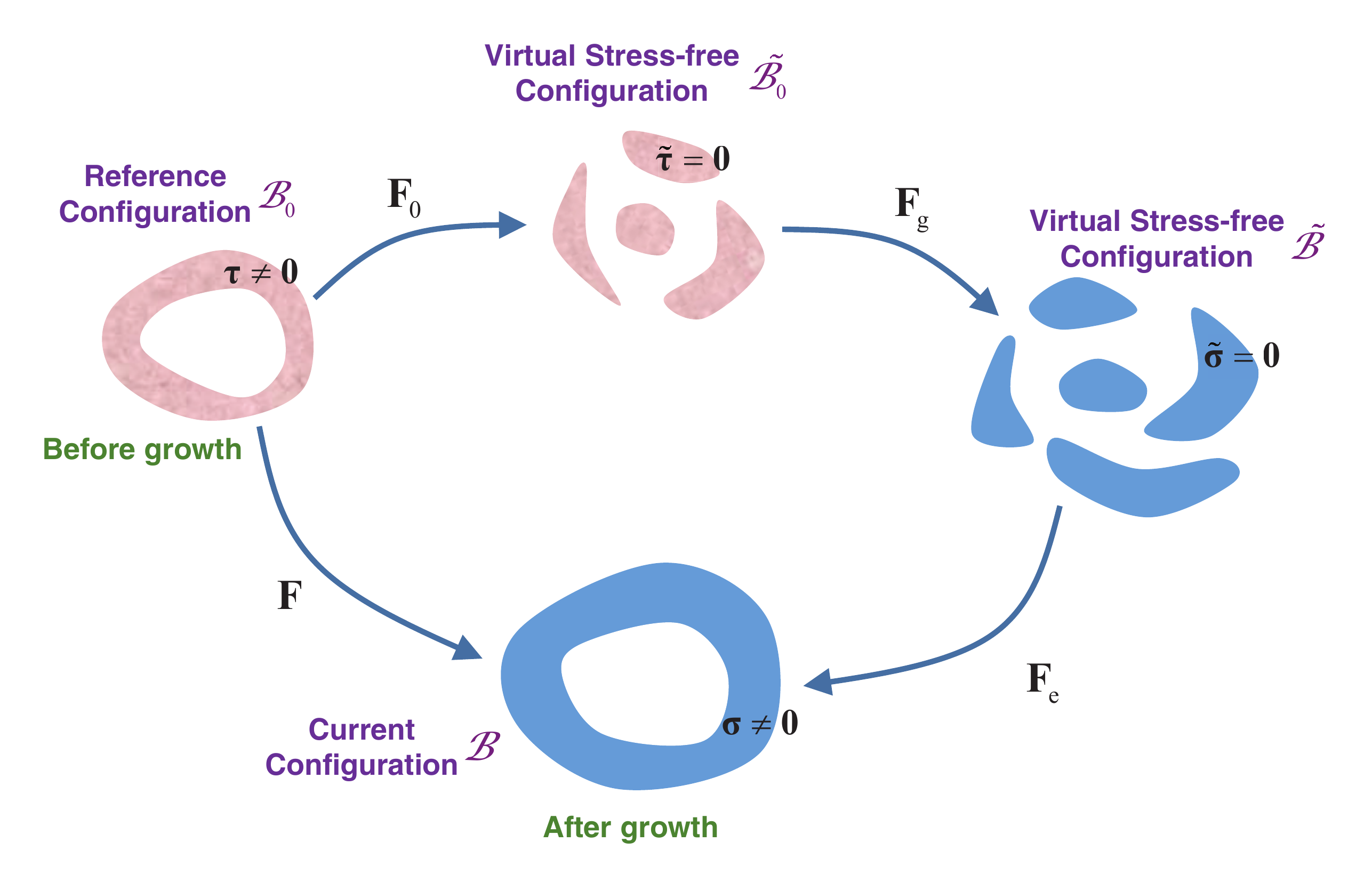}
	\caption{Diagram of the modified multiplicative decomposition method for initially stressed, growing materials. }
	\label{Fig:3}
\end{figure}

With this consideration in mind, we propose to modify the reference configuration ${{\mathcal{B}}_{0}}$ of the growing tissue by regarding it as a state endowed with a certain level of residual stress $\bm{\tau}$ ($\bm{\tau}\ne \mathbf{0}$). 
\color{black}
Here, we are not concerned with the origin of the initial stress $\bm{\tau}$, and the boundary conditions in the initial reference configuration may be arbitrarily in an unloaded or a loaded state.
\color{black}
Starting from this state, the tissue grows to the current configuration $\mathcal{B}$ through which a total deformation $\mathbf{F}$ occurs (Figure 3). 
In order to determine quantitatively the total deformation $\mathbf{F}$ due to growth, the multiplicative decomposition method \citep{RN101} is adopted with modifications.  Since the initial state or reference configuration is modified to include an initial residual stress $\bm{\tau}$, our decomposition is modified by adding a step to release the initial residual stress so that the continuous tissue is discretized into an infinite number of elements that are free of stress, i.e. the virtual stress-free configuration ${{\tilde{\mathcal{B}}}_{0}}$.  
Afterwards, the tissue grows (or shrinks) to a state with more (or less) mass or volume but without residual stress, i.e. another virtual stress-free configuration $\tilde{\mathcal{B}}$. 
Then, the discrete elements are integrated into the final continuous body in the current configuration $\mathcal{B}$ (Figure \ref{Fig:3}). 
Here, the second and third steps are similar to those used in the conventional MD model \citep{RN101}, and the main difference is in the first step.

Similarly, we assume that unconstrained growth only occurs between two stress-free states, while releasing residual stress can only induce elastic deformations or the residual stress is only created by elastic deformations (Figure \ref{Fig:3}). 
Therefore, the total deformation may be expressed as
\begin{equation} \label{F}
\mathbf{F} = {{\mathbf{F}}_{{e}}}{{\mathbf{F}}_{{g}}}{{\mathbf{F}}_{0}},
\end{equation}
where ${{\mathbf{F}}_{{e}}}$ is the elastic deformation, ${{\mathbf{F}}_{{g}}}$ is the growth deformation, and ${{\mathbf{F}}_{{0}}}$ induced by releasing the initial residual stress from the body. 
This latter deformation gradient refers to what we call the \textit{initial elastic deformation}. 
 
For this modified multiplicative decomposition (MMD) model, there are still some unsolved questions since the initial stress and virtual stress-free configuration were introduced. For instance, what is the form of the free energy function when we simultaneously consider initial stress and growth factor? For a specific real living matter, how can we obtain the initial stress distribution? Moreover, for a given initial stress distribution, how can we construct the relationship between the initial elastic deformation and the initial stress? 

\color{black}
First, for a pure elastic deformation in a continuum, the strain energy density can be defined as a function of the elastic deformation gradient tensor. So, from the virtual stress-free configuration $\tilde{\mathcal{B}}$ to the current configuration $\mathcal{B}$, the energy density function for stress-free materials can be defined as $\psi =\psi \left( {{\mathbf{F}}_{{e}}} \right)$. In addition, based on the decomposition shown in Figure 3 and the Eq. (1), the elastic deformation required to create residual stress is  ${{\mathbf{F}}_{{e}}}=\mathbf{FF}_{0}^{-1}\mathbf{F}_{{g}}^{-1}$. Then, for this current, initially stressed, growing matter, the free energy function can be expressed as
\begin{equation}
\psi \left( {{\mathbf{F}}_{{e}}} \right)=\psi \left( \mathbf{FF}_{0}^{-1}\mathbf{F}_{{g}}^{-1} \right),
\end{equation}
where the growth deformation gradient tensor   is assumed to be independent of the stress states \citep{ RN90}.
\color{black}

Second, we recall that there are some existing methods to access the distribution of initial residual stresses beyond the destructive experiments such as the opening angle method. For instance, with the minimal stress gradient method combining the initial stress symmetry (ISS) condition for the initially stressed materials, \cite{ RN389} showed that the initial stress distribution can be accessed via the Cauchy stress distribution solved by the minimal stress gradient method in the current configuration. 
Alternatively, by adopting the Airy stress function method, \cite{ RN487} gave three distribution forms of residual stresses satisfying the equilibrium equation and proposed a morphological method to quantify their magnitude. 

Finally, by assuming that the constitutive equation for the stress-free material is known and invertible, we will show that the initial elastic deformation for incompressible materials can be solved via a method proposed by \cite{ RN483} .


\subsection{The virtual stress-free configuration}


The most basic assumption in this current MMD decomposition is that the stress-free state is a discrete configuration and is unavailable in practice, which means there is no real continuum configuration and it is difficult to define the deformation. Here, we use the limitation concept proposed by \cite{RN483} to access approximately the stress-free configuration by letting the infinitesimal volume surrounding a material point tend to zero. This derivation is in the same spirits with the proof of Cauchy's theorem and we do not reproduce it here to save space. In short, it shows that the real stress-free configuration is made of infinitesimal discrete regions and that the elastic deformation gradient tensor ${{\mathbf{F}}_{0}}$ for a stress-free material can be mathematically approximated to the corresponding elastic deformation gradient tensor for a continuum material. Therefore, the usual constitutive equations of continuum mechanics such as the neo-Hookean, Mooney-Rivlin, or Fung models can be used for the stress-free materials.


\subsection{Governing equations} 
\label{section2.3}

Here, if we consider the internal material constrains, the free energy function will be modified by $\psi \rightarrow \psi - pC$, where $p$ is a Lagrange multiplier,  and $C$ is a scalar function encapsulating the internal elastic  constraints. Then the nominal stress $\mathbf{S}$ is obtained as
\begin{equation}
\mathbf{S}=J\left( \frac{\partial \psi }{\partial \mathbf{F}}-p\frac{\partial C}{\partial \mathbf{F}} \right)=J\mathbf{F}_{g{0}}^{-1}\left( \frac{\partial \psi }{\partial {{\mathbf{F}}_{e}}}-p\frac{\partial C}{\partial \mathbf{F}_e} \right). 
\end{equation}
\color{black}
where ${{\mathbf{F}}_{{g0}}}={{\mathbf{F}}_{{g}}}{{\mathbf{F}}_{0}}$, and the volume change $J=\det \left( \mathbf{F} \right)$ appears because the elastic strains are computed from the grown state. 
\color{black}
In particular, for elastic \textit{incompressible} materials, ${C}=\det \left( {{\mathbf{F}}_{{e}}} \right)-{1}$. Then the nominal stress for incompressible materials becomes
\begin{equation}
\mathbf{S}=J\left( \frac{\partial \psi }{\partial \mathbf{F}}-p{{\mathbf{F}}^{-1}} \right)=J\mathbf{F}_{g{0}}^{-1}\left( \frac{\partial \psi }{\partial {{\mathbf{F}}_{e}}}-p\mathbf{F}_{e}^{-1} \right).
\end{equation}
Because the relationship between the Cauchy stress $\bm{\sigma}$ and the nominal stress $\mathbf{S}$ is $\bm{\sigma}={{J}^{-1}}\mathbf{FS}$, we have in general,
\begin{equation}
\bm{\sigma}={{J}^{-1}}{{\mathbf{F}}_{e}}{{\mathbf{F}}_{g{0}}}J\mathbf{F}_{g{0}}^{-1}\left( \frac{\partial \psi }{\partial {{\mathbf{F}}_{e}}}-p\frac{\partial C}{\partial {{\mathbf{F}}_{e}}} \right)={{\mathbf{F}}_{e}}\left( \frac{\partial \psi }{\partial {{\mathbf{F}}_{e}}}-p\frac{\partial C}{\partial {{\mathbf{F}}_{e}}} \right),
\end{equation}
and for elastic incompressible materials,
\begin{equation}
\bm{\sigma}={{\mathbf{F}}_{e}}\left( \frac{\partial \psi }{\partial {{\mathbf{F}}_{e}}}-p\mathbf{F}_{e}^{-1} \right).
\end{equation}
In the absence of body forces, the equation of equilibrium are
\begin{equation}
\text{Div }\mathbf{S}=\mathbf{0},\qquad \text{div }\bm{\sigma}=\mathbf{0},
\end{equation}
where the Div and div are divergence operators in ${{\mathcal{B}}_{{0}}}$ and $\mathcal{B}$, respectively, and the boundary conditions are
\begin{equation}
{{\mathbf{S}}^{\text{T}}}\mathbf{N}=\mathbf{0},\qquad {{\bm{\sigma}}^{\text{T}}}\mathbf{n}=\mathbf{0}.
\end{equation}


\subsection{Key issues for initially stressed growing matter} 
\label{section2.4}


So far we have presented the kinematic description, the basic assumptions, and the governing equations of the modified multiplicative decomposition (MMD) growth model for initially stressed biological tissues. However, we still need to clarify some key issues: (1) selection of energy density function; (2) determination of initial elastic deformation; (3) determination of initial residual stress.


\subsubsection{Selection of free energy density function} 

\color{black}For materials with the initial residual stress $\bm{\tau}$, the cauchy stress  $\bm{\sigma}$ in the current configuration satisfies the constitutive equation $\bm{\sigma}=\bm{\zeta }\left( \mathbf{F},\bm{\tau} \right)$.Here the free energy density function yielding $\bm{\sigma}=\bm{\zeta }\left( \mathbf{F},\bm{\tau} \right)$  should also satisfy the requirement of initial stress symmetry (ISS) in \cite{RN389,Gower17}. According to ISS, when we exchange the roles of the current configuration and the reference configuration, the initial stress needs to satisfy the equation  $\bm{\tau} =\bm{\zeta}\left( {{\mathbf{F}}^{-1}},\bm{\sigma} \right)$, showing that the constitutive function has no preferred reference configuration. When $\mathbf{F}=\mathbf{I}$, we obtain $\bm{\tau}=\bm{\zeta}\left( \mathbf{I}{,}\bm{\tau} \right)$ which indicates that ISS can also recover the initial stress compatibility (ISC). As pointed out by Gower et al. (2017), the restrictions of ISS are a consequence of energy conservation rather an assumption made for convenience. These restrictions ensure that the predicted stress and strain energy do not depend upon an arbitrary choice of reference configuration. If a strain-energy function does not satisfy these restrictions, the resulting constitutive response function may lead to unphysical behavior \citep{Gower17}.\color{black}

Moreover, \cite{ RN389} also showed that the constitutive equation of an initially stressed material derived by using a virtual stress-free configuration satisfies ISS. 
\color{black}
For an initially stressed growing material, we now explore the consequences of the swapping of the reference configuration and the current configuration for the constitutive equations and show the constitutive equation that satisfy the ISS condition.
\color{black}

We call $\bm\vartheta$ the constitutive law of the body, giving the stress for a deformation taking place in an initially stress-free state. Both $\bm{\tau}$ and $\bm{\sigma}$ are related to their respective stress-free configuration ${{\tilde{\mathcal{B}}}_{{0}}}$ and $\tilde{\mathcal{B}}$ by the same constitutive law, $\bm\vartheta$. 
Explicitly, we see from Figure \ref{Fig:3} that
\begin{equation} \label{tau}
	\bm{\tau}=\bm{\vartheta} \left( \mathbf{F}_{0}^{-1}\mathbf{F}_{{0}}^{-\text{T}},{{p}_{0}} \right),
	\qquad \bm{\sigma}=\bm{\vartheta} \left( {{\mathbf{F}}_{{e}}}\mathbf{F}_{{e}}^{\text{T}},p \right),
\end{equation}
where the scalars ${{p}_{0}}$ and $p$ are arbitrary Lagrange multipliers, to be determined from the boundary conditions on $\partial {{\mathcal{B}}_{{0}}}$ and $\partial \mathcal{B}$, respectively. 
Now by Eq.\eqref{F}, ${{\mathbf{F}}_{{e}}}=\mathbf{FF}_{0}^{-1}\mathbf{F}_{{g}}^{-1}$, and Eq.\eqref{tau}$_2$ can be rewritten as 
\begin{equation} \label{theta}
	\bm{\sigma}=\bm{\vartheta} \left( \mathbf{FF}_{0}^{-1}\mathbf{F}_{{g}}^{-1}\mathbf{F}_{{g}}^{-\text{T}}\mathbf{F}_{0}^{-\text{T}}{{\mathbf{F}}^{\text{T}}},p \right)=\bm{\zeta} \left( \mathbf{F},{{\mathbf{F}}_{g}},\bm{\tau},p \right),
\end{equation}
where $\mathbf{\bm{\zeta} }$ is the constitutive equation for the \textit{initially-stressed, growing} materials. Then, we may swap the configuration ${{\mathcal{B}}_{{0}}}$ and $\mathcal{B}$ by performing the following swaps for the fields, see Figure \ref{Fig:3},
\begin{equation}
	\mathbf{F} \rightarrow {{\mathbf{F}}^{-1}},\quad  {{\mathbf{F}}_{{g}}} \rightarrow \mathbf{F}_{{g}}^{-1},\quad
	 {{\mathbf{F}}_{{0}}} \rightarrow \mathbf{F}_{{e}}^{-1},\quad {{\mathbf{F}}_{{e}}}\rightarrow \mathbf{F}_{{0}}^{-1},\quad {{p}_{0}}\rightarrow p.
\end{equation}
With these swaps, Equation \eqref{tau}$_{2}$ now reads as: $\bm{\tau}=\bm{\vartheta} \left( \mathbf{F}_{0}^{-1}\mathbf{F}_{{0}}^{-\text{T}},{{p}_{0}} \right)$ and Equation \eqref{tau}$_{1}$ now reads as: $\bm{\sigma}=\bm{\vartheta} \left( {{\mathbf{F}}_{{e}}}\mathbf{F}_{{e}}^{\text{T}},p \right)$. Effectively, $\bm{\tau}$ and $\bm{\sigma}$ have swapped roles, as required. However, this swapping has consequences on the constitutive law $\mathbf{\bm{\zeta} }$, because Eq.\eqref{theta} now reads
\begin{equation} \label{vartheta}
\bm{\vartheta} \left( {{\mathbf{F}}^{-1}}{{\mathbf{F}}_{{e}}}{{\mathbf{F}}_{{g}}}\mathbf{F}_{{g}}^{\text{T}}\mathbf{F}_{{e}}^{\text{T}}{{\mathbf{F}}^{-\text{T}}},{{p}_{0}} \right)=\bm{\zeta} \left( {{\mathbf{F}}^{-1}},\mathbf{F}_{{g}}^{-1},\bm{\sigma},{{p}_{0}} \right).
\end{equation}
Combining Eqs. \eqref{theta} and \eqref{vartheta} now shows that the initially stressed growing material satisfies ISS when a constitutive law of the form $\bm{\sigma}=\mathbf{\bm{\zeta} }\left( \mathbf{F},{{\mathbf{F}}_{{g}}},\bm{\tau},p \right)$ implies that $\bm{\tau}=\mathbf{\bm{\zeta} }\left( {{\mathbf{F}}^{-1}},\mathbf{F}_{{g}}^{-1},\bm{\sigma},\color{black}p_0\color{black} \right)$.
When a constitutive law is proposed without reference to virtual stress-free configuration, assuming ISS imposes restrictions on its form (see \cite{ RN389}  for these restrictions in a non-growing material).

With the introduction of virtual stress-free configurations, we can construct a constitutive law which satisfies ISS, and then we may work with that law without making any further reference to virtual stress-free configuration.

In Section \ref{section3} we present such a constitutive law, based on the neo-Hookean form of nonlinear elasticity. 
It presents the further advantage that it allows for the same residual stress to result from different pre-deformation decompositions, see \cite{Gower17}.
Note, however, that our analysis is not restricted to this constitutive choice and can easily be applied to other forms, such as the ones proposed by \cite{RN452} for instance.


\subsubsection{Determination of initial elastic deformation} 


In Section \ref{section2.3}, the constitutive equations for a growing material with initial residual stresses are expressed in terms of the initial stress $\bm{\tau}$, and not explicitly in terms of ${{\mathbf{F}}_{0}}$. 
Recalling the fundamental assumption that the constitutive equation for a stress-free material is unique and  invertible,  there must exist a relationship between ${{\mathbf{F}}_{0}}$ and $\bm{\tau}$. 
For infinitesimal deformation, \cite{ RN588} proposed an optimization algorithm to approximately obtain the initial strain component related to the initial pre-stress for layered piezoelectric structures. 
For finite deformation in soft material or biological tissues, \cite{ RN508, RN483} presented a method to access the left Cauchy- Green strain tensor ${{\mathbf{B}}_{0}}$ related to initial residual stress $\bm{\tau}$. 
However, note that it is impossible in general to obtain the uniquely explicit expression between the ${{\mathbf{F}}_{0}}$ and $\bm{\tau}$ \citep{ RN589}. 
Nevertheless, by assuming that growth deformation takes place along an axi-symmetric or principal direction, we show an example in the following Section 3 that the Cauchy stress can be obtained from a relation between ${{\mathbf{B}}_{0}}$ and $\bm{\tau}$.

Here we present the method proposed by \cite{ RN508, RN483} for obtaining the left Cauchy-Green tensor $\mathbf{B}_{0}^{(-1)}=\mathbf{F}_{0}^{-1}\mathbf{F}_{0}^{-\text{T}}$ from a given initial stress for growing elastic incompressible materials. Here, we note that $\mathbf{B}_{0}^{(-1)}$ is the left Cauchy-Green tensor of $\mathbf{F}_{0}^{-1}$ and does not equal to $\mathbf{B}_{0}^{-1}$
 
Based on Eq. ${(9)_1}$ and the assumption that the constitutive equation for stress-free materials is known and invertible, the left Cauchy-Green tensor $\mathbf{B}_{0}^{(-1)}$ for the deformation from ${{\tilde{\mathcal{B}}}_{{0}}}$ to ${{\mathcal{B}}_{{0}}}$ has a formal functional relationship to the initial stress $\bm{\tau}$ and the Lagrange multiplier ${p}_{0}$, written as
\begin{equation}
\mathbf{B}_{0}^{(-1)}=\hat{\bm{\vartheta} }\left( \bm{\tau},{{p}_{0}} \right).
\label{B0}
\end{equation}
From incompressibility $\det \left( \mathbf{F}_{0}^{-1} \right)=1$, it follows that
\begin{equation}
\det \mathbf{B}_{0}^{(-1)} = \det \left( \hat{\bm{\vartheta} }\left( \bm{\tau},{{p}_{0}} \right) \right)=1.
\label{detB0}
\end{equation}
By solving  Eq.\eqref{detB0}, ${{p}_{0}}$ can be obtained (at least, in principle). 
Then $\mathbf{B}_{0}^{(-1)}$ can also be computed by substituting ${{p}_{0}}$ and the given $\bm{\tau}$ into Eq.\eqref{B0}. 


\subsubsection{Initial residual stress distribution} 


There are two major ways to measure experimentally residual stresses. One relies on relaxation measurement methods, which are suitable to measure residual stresses in simple or axisymmetric shapes such as the tubular organs examined by the opening angle method. The other uses diffraction methods, including ultrasonic, photoelastic, and X-ray diffraction (Schajer, 2013).

From a theoretical point of view, it is difficult to access the distribution of initial stress, especially for biological tissue with an arbitrary shape or with a complex structure. For some simple or axisymmetric shapes, as found for tubular organs, we may use the stress potential function method \citep{RN384}, see Section \ref{section4}. 

\section{Growth of an initially stressed neo-Hookean tissue}
\label{section3}


Accounting now for the large deformation generated during the growth process, we present a simple constitutive equation for initially stressed growing materials based on the elastic constitutive equation of neo-Hookean solids. 


\subsection{Constitutive equation}


From the virtual stress-free configuration $\tilde{\mathcal{B}}$ to the current residually stressed configuration $\mathcal{B}$, the free energy density of a neo-Hookean solid is
\begin{equation}
\psi =\frac{\mu }{2}\left[ \text{tr}\left(\mathbf{F}_{e}^{\text{T}}{{\mathbf{F}}_{e}}\right) - 3 \right],
\label{neo}
\end{equation}
where $\mu $ is the initial shear modulus. Here we assume that the initial shear modulus remains constant during the growth process. 

Based on Eq.\eqref{F}, we have
the nominal stress $\mathbf{S}$ is
\begin{equation}
\mathbf{S}\text{=}J\left( \mu \mathbf{C}_{g\text{0}}^{-1} {{\mathbf{F}}^{\text{T}}}-p{{\mathbf{F}}^{-1}} \right),
\end{equation}
and the corresponding Cauchy stress $\bm{\sigma}$ is 
\begin{equation}
\bm{\sigma}=\mu \mathbf{F} \mathbf{C}_{\text{g0}}^{-1}{{\mathbf{F}}^{\text{T}}}-p\mathbf{I}.
\end{equation}

For \emph{axisymmetric growth deformation}, the deformation gradient tensors and the growth process are diagonal in their respective bases of orthogonal unit vectors. Then, by commutativity the Cauchy stress can be expressed as 
\begin{equation} \label{sigma-B0}
\bm{\sigma}=\mu \mathbf{B} \mathbf{B}_{g}^{-\text{1}}\mathbf{B}_{0}^{(-1)}-p\mathbf{I},	
\end{equation}
where $\mathbf B = \mathbf{FF}^\text{T}$, $\mathbf B_g = \mathbf{F}_g \mathbf F_g^\text{T}$, $\mathbf B_0^{(-1)} = \mathbf{F}_0^{-1}  \mathbf {F}_0^{-\text{T}}$.


\subsection{ The initial elastic deformation}


From Eq.\eqref{sigma-B0} written in ${{\mathcal{B}}_{\text{0}}}$ we see that  the initial residual stress $\bm{\tau}$ can be expressed as
\begin{equation}
\bm{\tau}=\mu \mathbf{B}_{0}^{(-1)}-{{p}_{0}}\mathbf{I},
\end{equation}
so that  $\mathbf{B}_{0}^{(-1)}$ can be found as
\begin{equation} \label{B0-1}
\mathbf{B}_{0}^{(-1)}=\frac{\bm{\tau}+{{p}_{0}}\mathbf{I}}{\mu }.
\end{equation}
Here, ${{p}_{0}}$ is the only yet unsolved parameter; it is related to the boundary condition in the initially stressed configuration. 

The first three principal invariants of $\mathbf{B}_{0}^{(-1)}$ are related to those of $\bm{\tau}$ as \citep{RN389}
\begin{equation} \label{IB0}
{{I}_{3,\mathbf{B}_{0}^{(-1)}}}=\frac{p_{0}^{3}+{{I}_{1,\bm{\tau}}}p_{0}^{2}\text{+}{{I}_{2,\bm{\tau}}}{{p}_{0}}\text{+}{{I}_{3,\bm{\tau}}}}{{{\mu }^{3}}}
\end{equation}
From the  incompressibility constraint, ${{I}_{3,\mathbf{B}_{0}^{(-1)}}} = 1$, so that Eq.\eqref{IB0} reads
\begin{equation} \label{p0}
p_{0}^{3}+{{I}_{1,\bm{\tau}}}p_{0}^{2}+{{I}_{2,\bm{\tau}}}{{p}_{0}}+{{I}_{3,\bm{\tau}}}-{{\mu }^{3}}=0.
\end{equation}
By solving Eq.\eqref{p0}, we obtain the Lagrange multiplier $p_0$ formally as ${{p}_{0}}=\wp (\bm{\tau})$. 
As the explicit form  is complicated, we do not present the corresponding explicit general expression here to save space.
\color{black}
It suffices to note that only one root of the cubic is relevant.
The details on how to identify the adequate root of Eq. \eqref{p0} are given by  \cite{RN389}, based on continuity of the root with changing residual stress. \color{black} 
Then the tensor $\mathbf{B}_{0}^{(-1)}$ can be written as 
\begin{equation}
\mathbf{B}_{0}^{(-1)}=\frac{\bm{\tau}+\wp (\bm{\tau})\mathbf{I}}{\mu }.
\end{equation}
Finally, we find the Cauchy stress for axisymmetric growth deformation based on Eq.\eqref{sigma-B0}.


\section{Growth stress of a tubular tissue}
\label{section4}


Tubular strutures such as plants, blood vessels, weasands, or gastro-intestinal walls are the most common biological tissues found in living organisms. 
Healthy organs always keep an ideal state, with moderate stress levels and a functional morphology. 
To understand further the growth or evolution rules for tubular organs, it is important to incorporate residual stress into growth theory. 
Also, it is vital to analyze the influence of the initial stress in reference configuration on the residual stress in the current configuration so that we can observe the growth process in a more real and practical way. 

\begin{figure}
	\centering
	\includegraphics[width=0.8\textwidth]{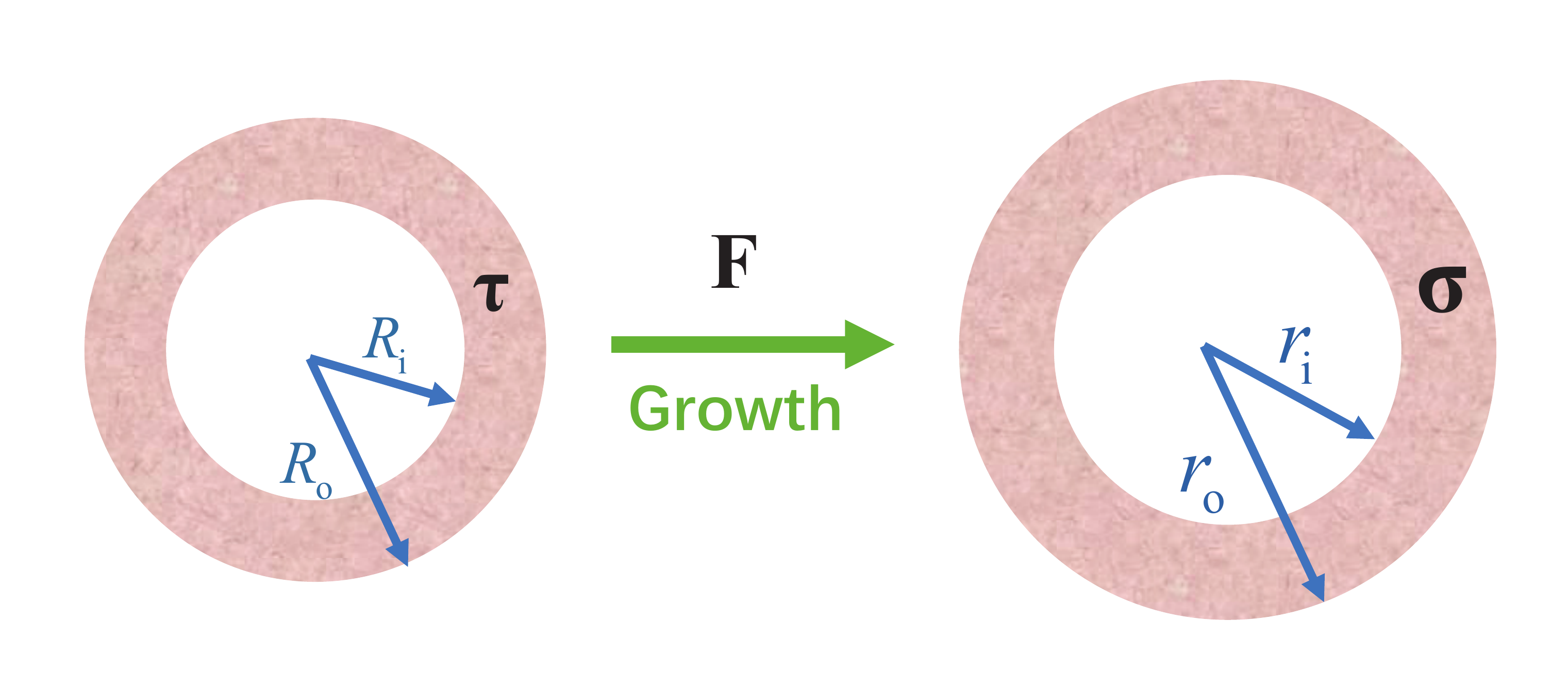}
	\caption{Growth of an initially stressed tube (plane strain model).}
	\label{Fig:4}
\end{figure}

Here we treat the example of a simplified plane strain growing axisymmetric tube model with an initial residual stress field $\bm{\tau}$, see Figure \ref{Fig:4}. 
The reference configuration ${{\mathcal{B}}_{\text{0}}}$ is associated with the cylindrical coordinates $\left( R,\mit{\Theta} ,Z \right)$, and the current configuration $\mathcal{B}$ with the coordinates $\left( r,\theta ,z \right)$.
\subsection{Distribution of the initial stress}
\color{black} 
We adopt the Airy stress function method to define a possible type of the distribution of the initial stress field  $\bm{\tau}$ which satisfies the self-equilibrium equation.
\color{black}

In the reference configuration ${{\mathcal{B}}_{\text{0}}}$, its non-zero components should satisfy
\begin{equation}
 \frac{\partial {{\tau }_{RR}}}{\partial R}+\frac{1}{R}\frac{\partial {{\tau }_{\mit{\Theta} R}}}{\partial \mit{\Theta} }+\frac{{{\tau }_{RR}}-{{\tau }_{\mit{\Theta \Theta}}}}{R}=0, \quad
  \frac{1}{R}\frac{\partial {{\tau }_{\mit{\Theta \Theta} }}}{\partial \mit{\Theta} }+\frac{\partial {{\tau }_{R\mit{\Theta }}}}{\partial  R }+\frac{{{\tau }_{\mit{\Theta} R}}+{{\tau }_{R \mit{\Theta} }}}{R}=0,
\end{equation}
subject to the traction-free boundary conditions $\tau_{RR} = \tau_{R\mit{\Theta}} = 0$ on the inner and outer surfaces at $R = R_i,R_o$. 
Then, introducing the Airy stress function $\phi \left( R,\mit{\Theta}  \right)$, we find that the general solution is
\begin{equation}
	{{\tau }_{RR}}=\frac{1}{R}{{\phi }_{,R}}+\frac{1}{{{R}^{2}}}{{\phi }_{,\mit{\Theta \Theta} }},\qquad 
	{{\tau }_{R\mit{\Theta}}}=\frac{1}{{{R}^{2}}}{{\phi }_{,\mit{\Theta} }}-\frac{1}{R}{{\phi }_{,\mit{\Theta} R}},\qquad
	{\tau }_{\mit{\Theta \Theta}}={{\phi }_{,RR}}.
\end{equation}
For solutions such that $\phi =\phi \left( R \right)$ only, this reduces to
\begin{equation}
	{{\tau }_{RR}}=\frac{1}{R}f(R),\qquad 
	{{\tau }_{R\mit{\Theta}}}=0, \qquad {\tau }_{\mit{\Theta \Theta}}={f}'(R),
\end{equation}
where $f = \phi'$ is a stress potential function. 

\begin{figure}
	\centering	
	\includegraphics[width=0.5\textwidth]{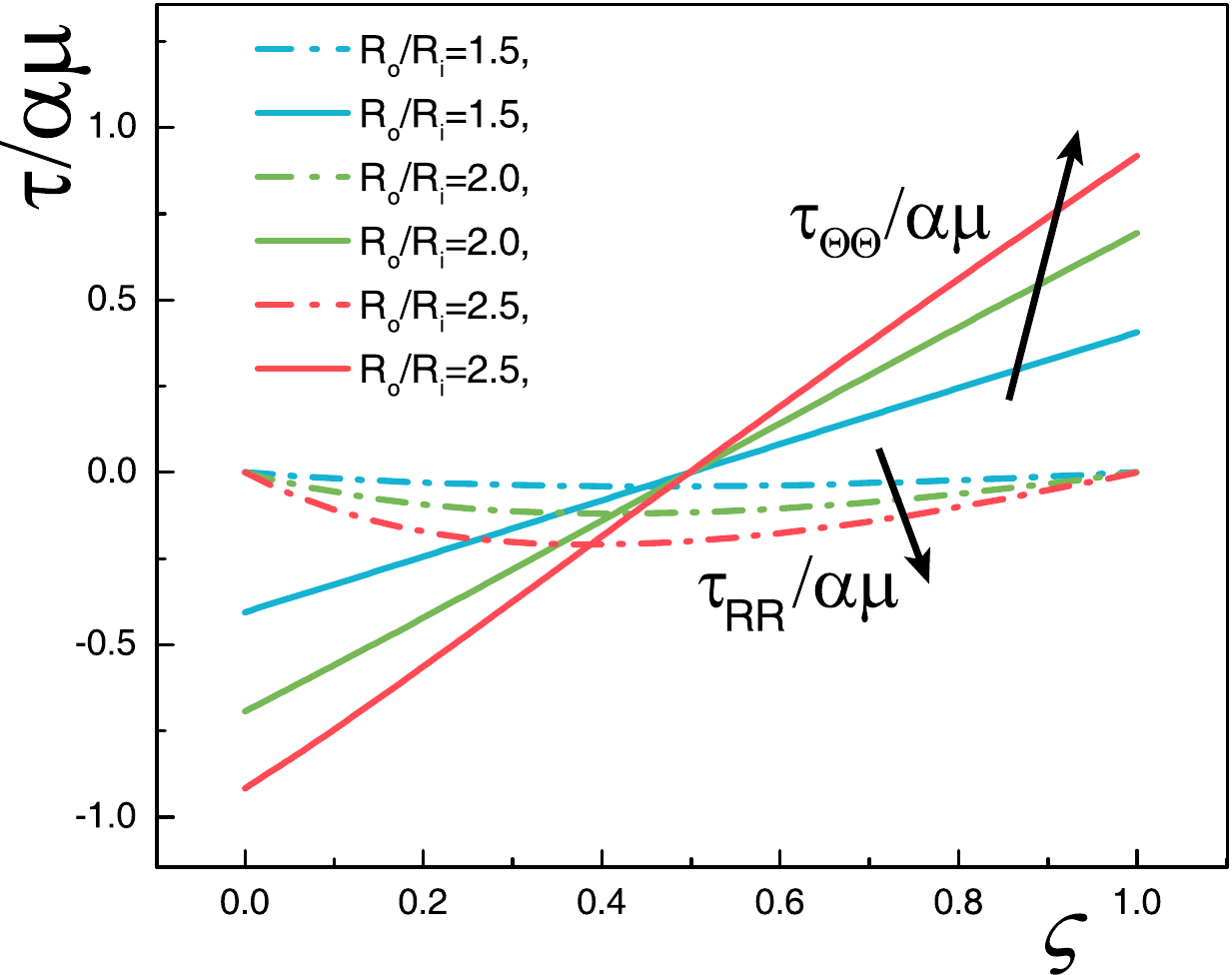}
	\caption{Distribution of initial stress (prior to growth) for different initial relative wall thicknesses ($R_o/R_i=1.5,2.0,2.5$), obtained from a logarithmic stress potential function. Here $\varsigma =\frac{R-{R_i}}{{{R}_{0}}-{R_i}}$ is the dimensionless radial coordinate.}
	\label{Fig:5}
\end{figure}

Here, we take a logarithmic stress potential function for illustration, 
\begin{equation}
	f(R)=\alpha \mu R\ln \left(R/R_i \right)\ln \left( R/R_o\right),
	\label{f}
\end{equation}
where $\alpha $ is a non-dimensional  measure of the residual stress amplitude. 
See \citep{RN384} for other examples of stress potential functions, such as parabolic and exponential variations. 
We also conducted the  analysis presented in this paper for those functions, and found similar results. 
For the logarithmic function, the radial stress component $\tau_{RR}$  varies almost linearly across the wall thickness, and is zero almost at the mid-thickness, see Figure \ref{Fig:5}.

The corresponding initial residual stress components are
\begin{equation}
\begin{split}
		& {{\tau }_{RR}}=\alpha \mu \ln \left( R/R_i\right)\ln \left( R/R_o\right), \\ 
		& {{\tau }_{\mit{\Theta \Theta}}}=\alpha \mu \ln \left( R/R_i\right)+\alpha \mu \ln \left( R/R_o\right)+\alpha \mu \ln \left( R/R_i\right)\ln \left( R/R_o\right). \\ 
\end{split}
\end{equation} 

Figure \ref{Fig:5} shows the resulting transmural distribution of the initial residual stress when $\alpha>0$. 
The radial stress is entirely tensile  ($\tau_{RR}<0$), is maximal at the middle thickness, and is small compared to the circumferential stress. 
The circumferential stress has an almost linearly antisymmetric variation with respect to the centroid surface, the maximum tensile stress is on the outer surface while the maximum compressive stress is on the inner surface (reversed when $\alpha <0$).


\subsection{Growth-induced residual stress}


Figure \ref{Fig:4}  and Figure \ref{Fig:3} show the overall diagram and the decomposition for the growth process. 
First, the corresponding deformation gradients read
\begin{equation}
	 \mathbf{F}=\text{diag}\left( \frac{\text{d}r}{\text{d}R},\ \frac{r}{R},\ 1 \right), \qquad
	{{\mathbf{F}}_{g}}=\text{diag}\left( {g_r},{{g}_{\theta }},1 \right),
\end{equation}
in their respective bases, where ${g_r}$, ${{g}_{\theta }}$ are the growth factors along the radial and circumferential directions, respectively. 

Incompressibility for the pure elastic deformation gradient tensor ${{\mathbf{F}}_{e}}=\mathbf{FF}_{0}^{-1}\mathbf{F}_{g}^{-1}$ reads as
$
\det \left( \mathbf{FF}_{0}^{-1}\mathbf{F}_{g}^{-1} \right)=1.
$
Because ${{\mathbf{F}}_{0}}$ is also assumed to correspond to a pure elastic deformation gradient, we have $\det \left( \mathbf{F}_{0}^{-1} \right)=1$, then the incompressibility condition reduces to 
$	\det \left( \mathbf{FF}_{g}^{-1} \right)=1$, which is integrated to
\begin{equation}
	{{r}^{2}}-r_{i}^{2}={g_r}{g_\theta}\left(R^2 - R_i^2 \right), \qquad \text{for } {R_i}\le R\le {R_o}.
\end{equation}
Then from Eqs.\eqref{sigma-B0} and \eqref{B0-1} we find the following non-zero Cauchy stress components,
\begin{equation}
 {{\sigma }_{rr}}=g_{r}^{-2}\left( {{\tau }_{RR}} + p_0 \right){{\left( \frac{\text{d}r}{\text{d}R} \right)}^{2}}-p, \qquad
 {{\sigma }_{\theta\theta}}=g_{\theta}^{-2}\left( {{\tau }_{\mit{\Theta \Theta} }} + p_0 \right){{\left( \frac{r}{R} \right)}^{2}}-p,
\end{equation}
and the sole non-zero equilibrium equation reads
\begin{equation} \label{eq}
r\frac{d {{\sigma }_{rr}}}{d r}+{{\sigma }_{rr}}-{{\sigma }_{\theta \theta }}=0,\qquad \text{for }\quad {{{r}}_{i}}<{r}<{{{r}}_{o}}.
\end{equation}

Next, we introduce the dimensionless initial radial position $\varsigma =\frac{R-{R_i}}{{{R}_{0}}-{R_i}}$ and Eq.\eqref{eq} becomes
\begin{multline}
\frac{d \sigma_{rr}}{d \varsigma} = \left[\left(\tau_{\mit{\Theta \Theta}} + p_0 \right) \left(\frac{r}{g_\theta R} \right)^\text{2} - \left(\tau_{RR} + p_0 \right) \left(\frac{g_\theta R}{r} \right)^\text{2} \right] 
\times \\[4pt]
\left[\frac{g_r g_\theta H \left(H \varsigma + R_i \right)}{g_r g_\theta H \varsigma \left(H \varsigma + 2 R_i \right) + r_i^2}\right],
\end{multline}
where $H={R_o}-{R_i}$ is the thickness of the tube in the initial stressed reference configuration. 
Integrating this expression subject to the boundary condition $\sigma_{rr}=0$ on the inner surface $\varsigma = 0$ ($r=r_i$), the growth-induced residual stress from an initial stress state is obtained as: 
\begin{multline} \label{sigma_rr}
 \sigma_{rr}(\varsigma) = \int_0^\varsigma \left[ \left(\tau_{\mit{\Theta \Theta}} + p_0\right) \left(\frac{r}{g_\theta R} \right)^2 - \left(\tau_{RR} + p_0 \right) \left(\frac{g_\theta R}{r} \right)^2 \right] \times
 \\[4pt]
\left[ \frac{{g_r}{g_\theta}H\left( \varsigma H+{R_i} \right)}{{g_r}{g_\theta}\varsigma H\left( \varsigma H+2{R_i} \right)+r_{i}^{2}} \right]d\varsigma.
\end{multline}
Finally, imposing the boundary condition $\sigma_{rr} =0$ on the outer surface at $\varsigma =1$ ($r = r_o$) in this expression, we access the value of the inner radius $r_i$ for a given initial stress, initial geometry and growth tensor. 
Then the Cauchy stress components follow from Eqs.\eqref{sigma_rr} and \eqref{eq}.


\subsection{Results}


First we take the current MMD growth model to start from an initial zero stress state, and check that we recover the results to the MD model \citep{RN101}, when $\alpha =0$ (no residual stress), ${g_r}=1$ (no radial growth), ${g_\theta}\ne 1$ (circumferential growth only). 
\begin{figure}
	\centering	
	\includegraphics[width=0.49\textwidth]{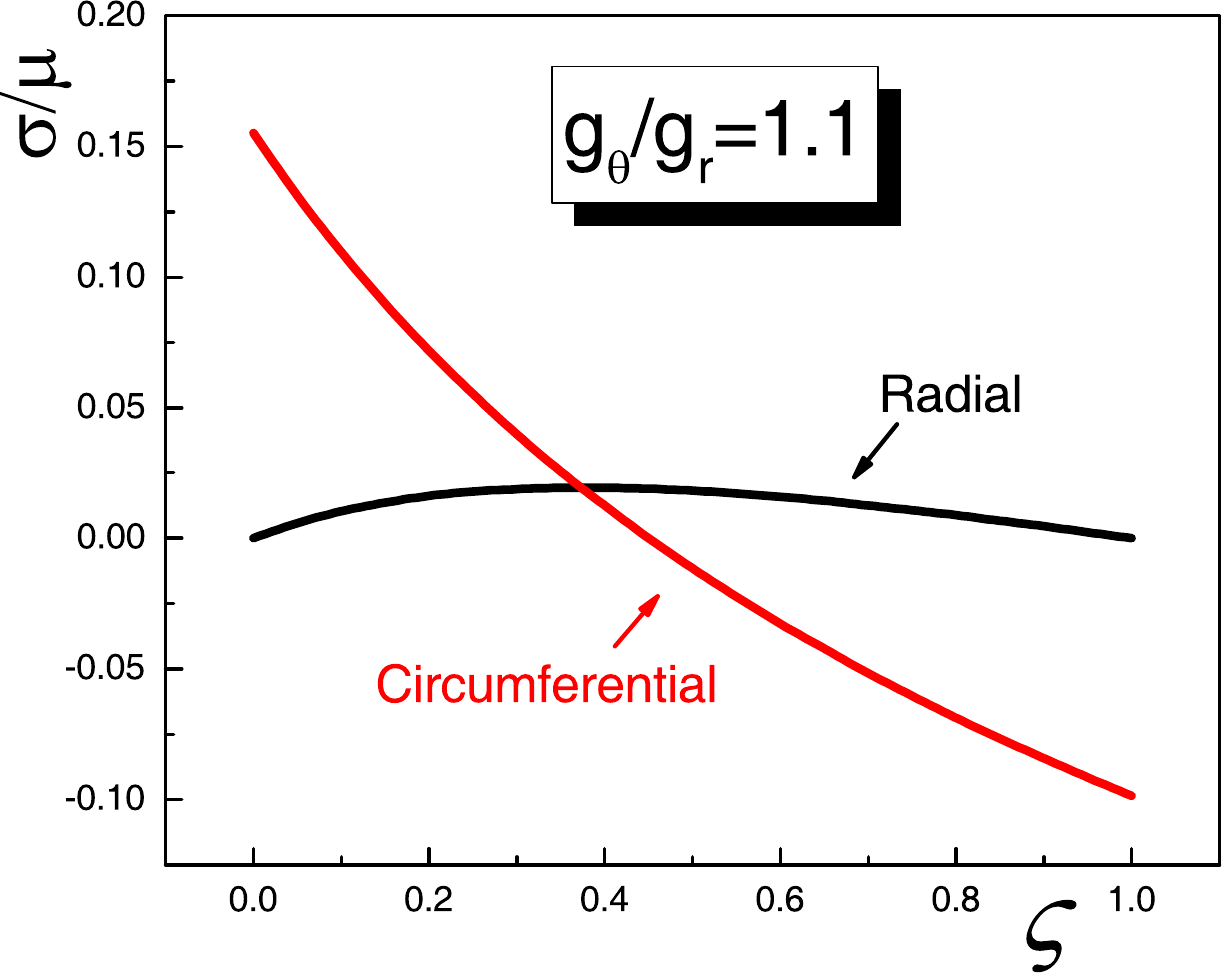}
	\includegraphics[width=0.49\textwidth]{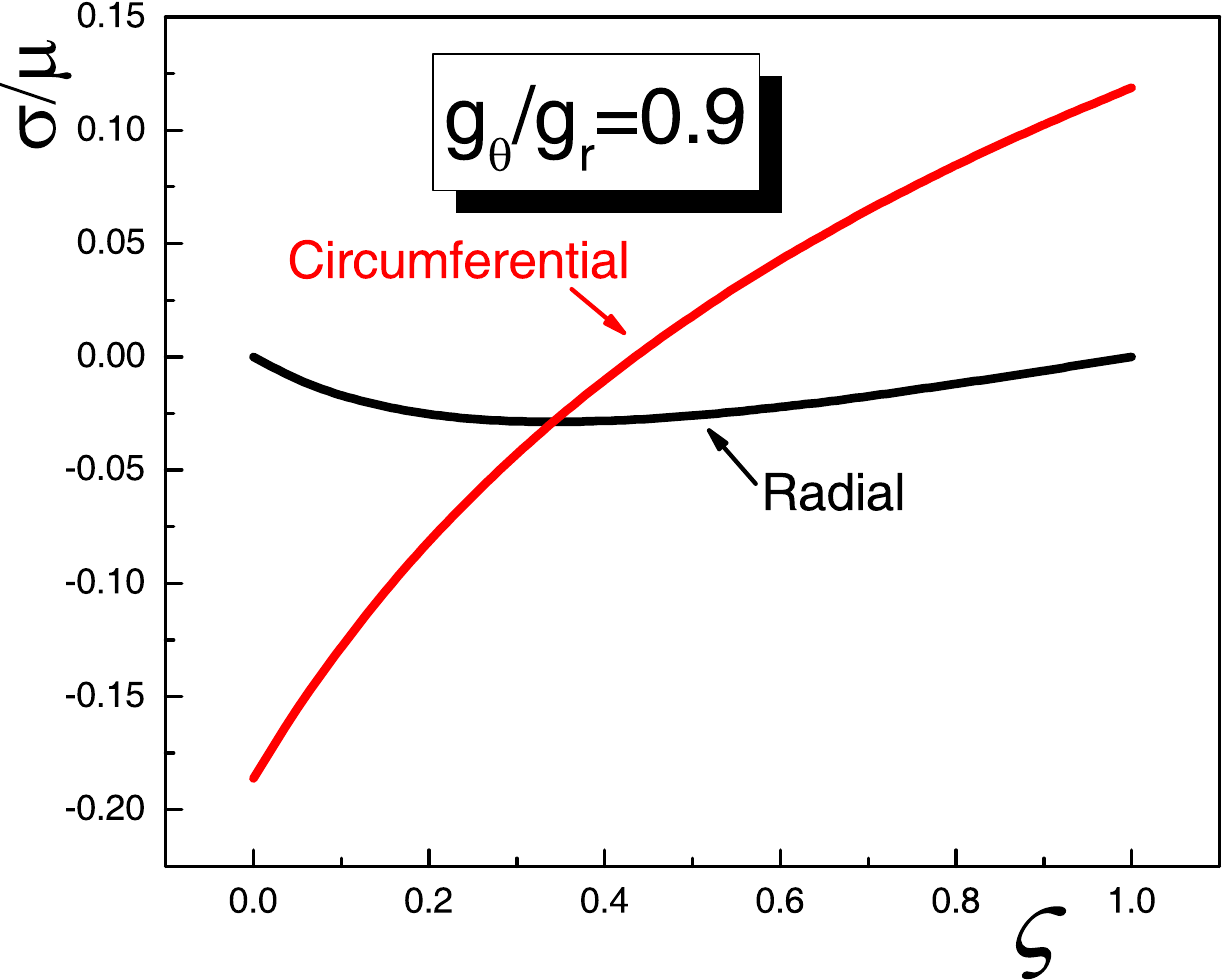}
	\caption{Transmural distributions of growth-induced residual stresses starting from an initial zero stress state (the classical multiplicative decomposition (MD) model of \cite{RN101}): circumferential growth only (left: $g_\theta = 1.1$, $g_r=1.0$; right: $g_\theta = 0.9$, $g_r=1.0$).}
	\label{Fig:6}
\end{figure}

Figure \ref{Fig:6} shows the resulting transmural distribution of differential growth-induced residual stress (``differential growth" means that the body grows differently along different directions). 
For circumferential expansion, the circumferential stress decreases monotonically from a tensile stress on the inner surface to compressive stress on the outer surface and this behavior is reversed for circumferential shrinkage.
As expected, we recover the results from \cite{RN101}.

Now we consider a more realistic scenario, where an initial residual stress exists in the reference configuration, and compare the results to those without considering the initial stress. 

\begin{figure}
	\centering	
	\includegraphics[width=0.49\textwidth]{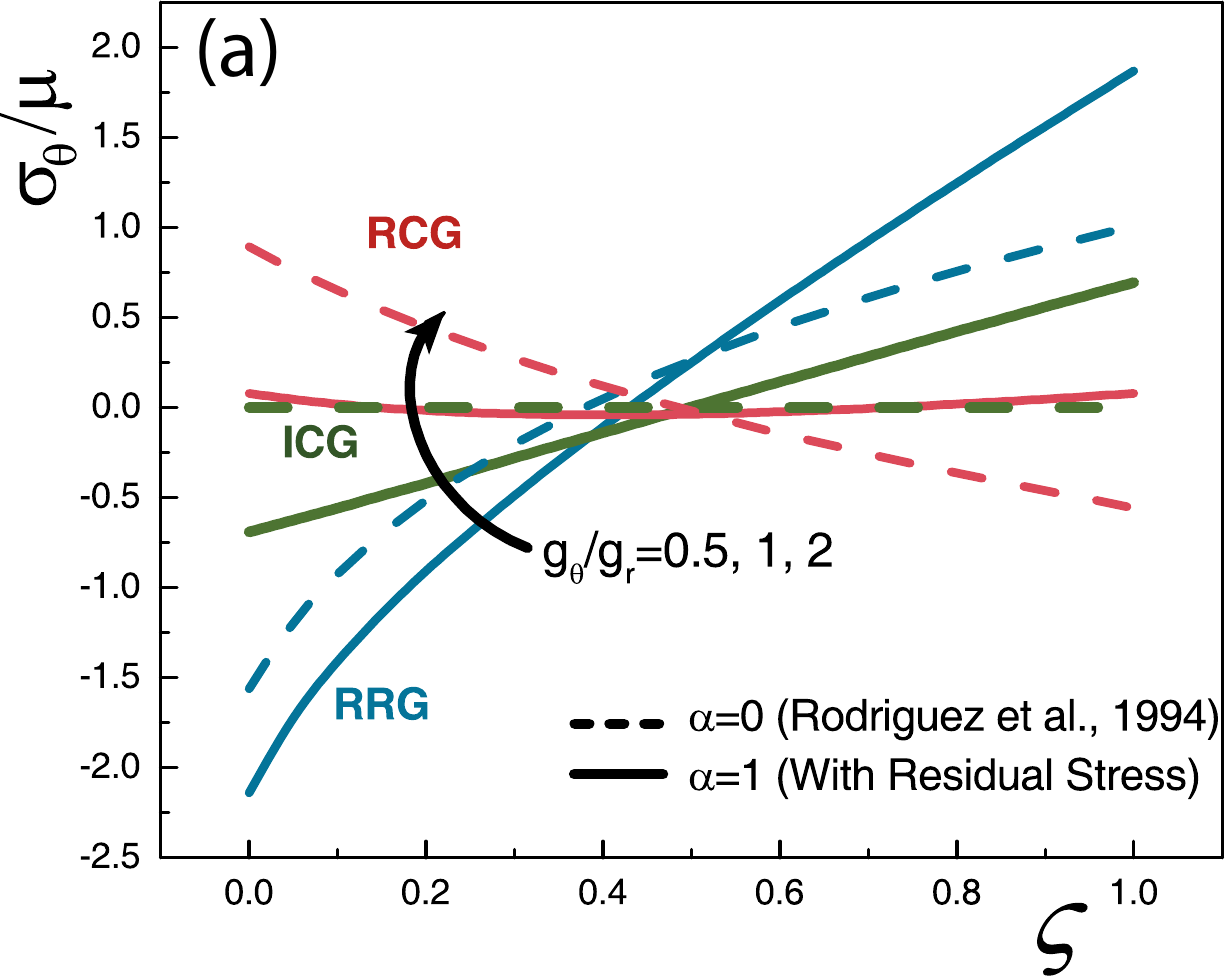} \
	\includegraphics[width=0.49\textwidth]{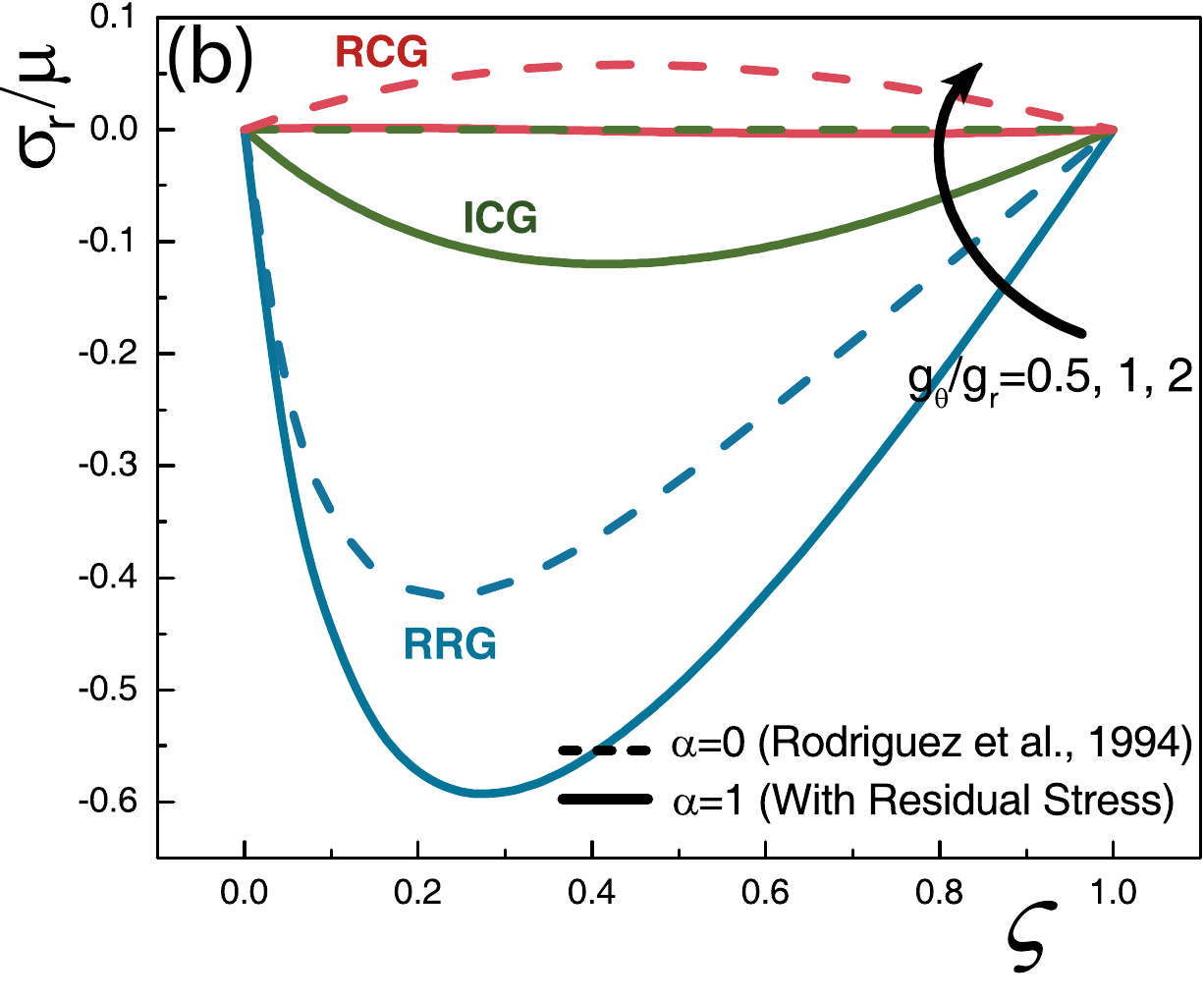}
	\caption{The transmural distribution of residual stresses for different differential growth ratios ($\alpha=1$, full lines): relative radial growth (RRG), when $g_\theta / g_r = 0.5$; isotopically compatible growth (ICG), when $g_\theta /g_r = 1.0$; and relative circumferential growth (RCG), when ${{g_\theta}}/{{g_r}}=2.0$, compared with Rodriguez et al.'s (1994) model (without initial residual stress, $\alpha=0$, dashed lines).}
	\label{Fig:7}
\end{figure}

Figure \ref{Fig:7} displays the comparative results: the solid lines correspond to the initial stress state with magnitude $\alpha =1.0$ and the dashed lines are for no initial stress ($\alpha =0$, as in Figure \ref{Fig:6}). 
The residual stresses depend on the ratio of growth factors ${{g_\theta}}/{{g_r}}$, and we study the following differential growth scenarios: relative radial growth (RRG), when $g_\theta / g_r < 1$; isotropically compatible growth (ICG), when $ g_\theta /g_r = 1$; and relative circumferential growth (RCG), when ${{g_\theta}}/{{g_r}}>1$. 
Figures \ref{Fig:7}(a) and (b) show the transmural distribution of circumferential stress and radial stress, respectively. 

For isotropically compatible growth (ICG, green curves), the initial residual stresses predicted by the MD model \citep{RN101} are zero throughout. 
By contrast, the MMD growth model with $\alpha =1.0$ is endowed with a significant distribution of initial residual stress. 
But of course, no matter what the magnitude of the isotropically compatible growth is as long as $g_\theta /g_r = 1.0$, the curves remain the same for both MD and MMD models, confirming that ICG does not give rise to growth-induced residual stress. 

Going now from ${{g_\theta}}/{{g_r}}={1.0}$ to  relative circumferential growth with ${{g_\theta}}/{{g_r}}= 2.0$ (RCG, red curves), we see that the distribution of growth-induced residual stress in circumferential direction produces an approximate clockwise rotation, both for the MD and MMD models. 
Specifically, on the inner side, the circumferential stress changes from zero stress to tensile stress in the MD model and changes from compressive stress to tensile stress in  the MMD model. 
Analogously, going from ${{g_\theta}}/{{g_r}}={1.0}$ to relative radial growth  with ${{g_\theta}}/{{g_r}}= 0.5$ (RRG, blue curves), we see an approximate anticlockwise rotation and an increasing value of compressive stress on the inner surface. 

\begin{figure}
	\centering	
	\includegraphics[width=0.49\textwidth]{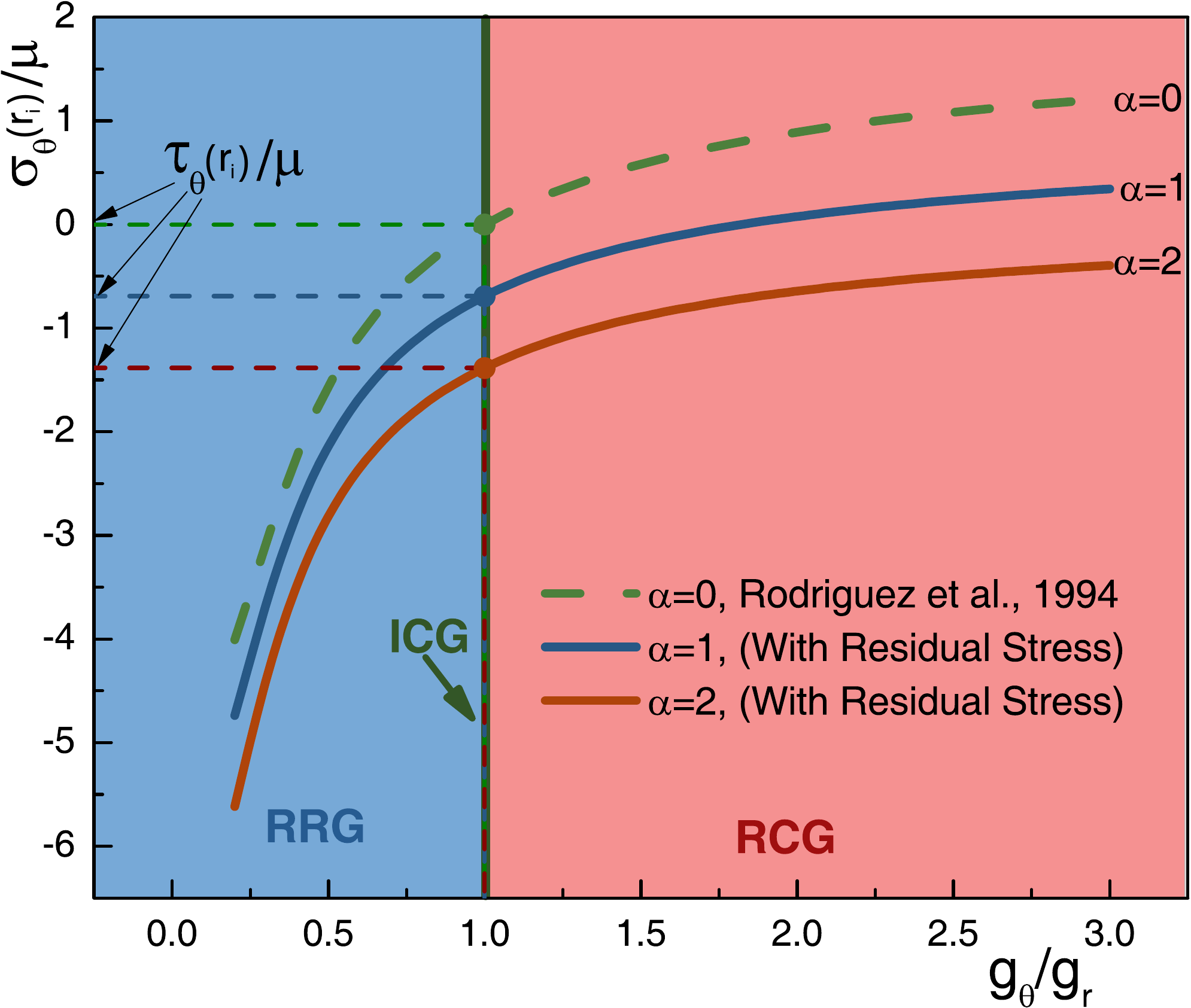}
	\includegraphics[width=0.49\textwidth]{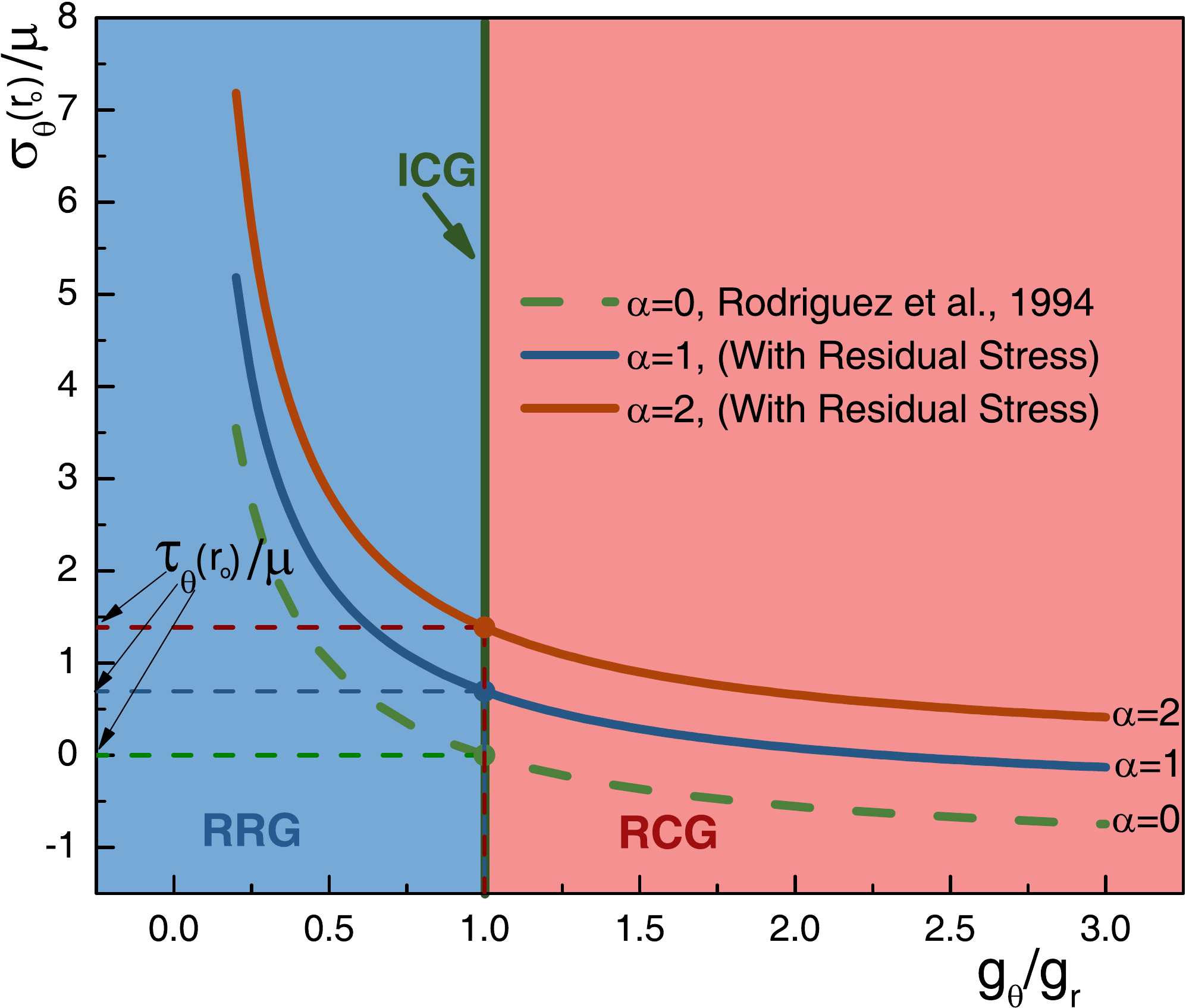}
	\caption{Variations of the residual circumferential stresses on the inner ($r  = r_i$) and outer ($r=r_o$) faces of the tube with the differential growth ratio (RRG: $g_\theta/g_r<1$, ICG: $g_\theta/g_r=1$, RCG: $g_\theta/g_r>1$), with ($\alpha=1.0, 2.0$, full lines) and without ($\alpha = 0$, dashed lines)  an initial residual stress \citep{RN101}.}
	\label{Fig:8}
\end{figure}

Figure \ref{Fig:8} shows the changes in circumferential stresses on the inner and outer surfaces with the differential growth ratio $g_\theta/g_r$. 
On the inner side, circumferential growth creates tensile stress while radial growth creates compressive stress, and \emph{vice-versa} on the outer side. 
So here, greater circumferential (radial) growth ratios lead to greater tensile (compressive) stresses and the introduction of initial residual stress accentuates these trends.
Clearly, the final distribution of residual stress depends not only on the differential growth ratio, but is also affected by the magnitude of initial stress. 

In the next section, we study the appearance of wrinkles due to loss of stability for a growing, initially stressed cylinder tube and investigate the role played by these factors on the development of its morphology.


\section{Growth-induced morphology of an initially stressed tube}
\label{section5}


Residual stress accumulates as a tubular organ grows. 
Similarly to tubes that develop circumferential instability under a critical pressure, leading to a non-circular cross-section \citep{RN571}, we expect our tube to buckle with increasing residual stress induced by the accumulation of differential growth. 
To understand the generation and the development of wrinkles on the inner side of the tube, we now conduct an instability analysis for the residually stressed state. 
Using linearized incremental theory, \cite{RN529}, \cite{RN90}, \cite{RN506}, \cite{RN511}, \cite{RN98}, \cite{RN505}, among others, have found the critical differential growth ratio leading to instability in a tube with no initial residual initial stress. 
Here, we use the MMD growth model to show the influence of an initial residual stress on the critical differential growth ratio and the resulting instability patterns.


\subsection{Incremental theory}

\begin{figure}
	\centering	
	\includegraphics[width=0.88\textwidth]{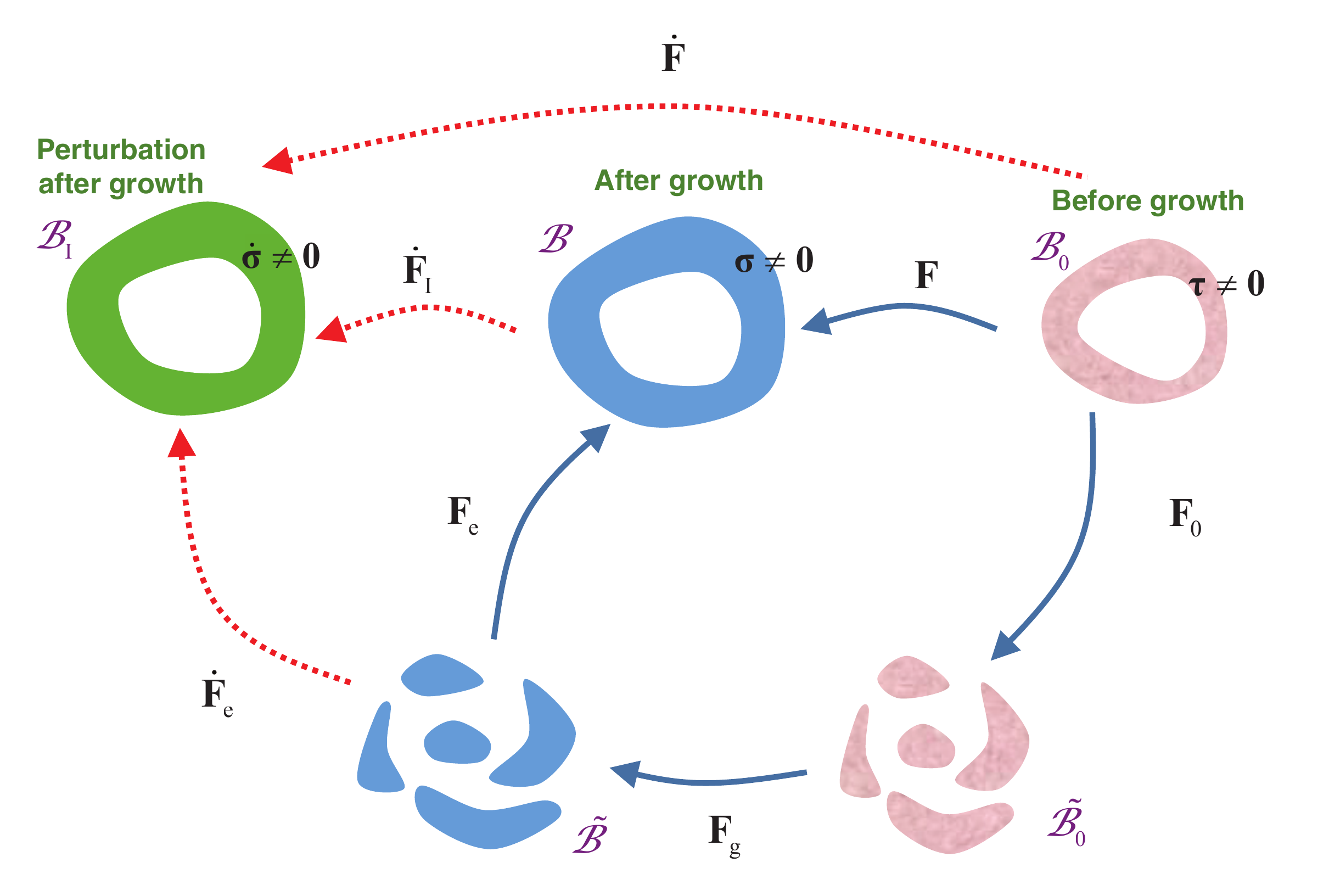}
	\caption{Kinematics of  incremental theory for the growth of initially stressed materials. 
		Full (red) lines show the displacement gradients and dashed (blue) lines the deformation gradient tensor.}
	\label{Fig:9}
\end{figure}
Following the growth process, an infinitesimal elastic deformation $\bm{\chi'}$ is applied in the current configuration $\mathcal{B}$ relative to the reference configuration $\mathcal{B}_0$, so that the particle position in the new configuration $\mathcal B_{\text{I}}$ can be expressed as $\mathbf{x'}  = \bm{\chi '} \left( \mathbf{X} \right)$.

Letting $\left(\text{d} \mathbf{x} \right)^{\bm{\cdot}} = \mathbf{x'} - \mathbf{x}$ denote the incremental displacement related to the reference configuration, we introduce the incremental displacement gradients $\mathbf{\dot{F}} = \left(\partial \mathbf{x}\right)^{\bm{\cdot}}/\partial\mathbf{X}$ with respect to the initial reference configuration ${{\mathcal{B}}_{0}}$, and $\mathbf{\dot{F}}_{\text{I}} = \left(\partial\mathbf{x} \right)^{\bm{\cdot}}/\partial\mathbf{x}$ with respect to the current configuration $\mathcal{B}$. 
By the chain rule, they are related to the deformation gradient tensor $\mathbf{F}={\partial\mathbf{x}}/\partial\mathbf{X}$ through 
\begin{equation} \label{Fdot}
	\mathbf{\dot{F}}={{\mathbf{\dot{F}}}_{\text{I}}}\mathbf{F}.
\end{equation}

Based on the incremental theory for tissue growth introduced by \cite{RN90}, we assume that the incremental deformation is infinitesimal and transient, so that the growth process is independent of the stress and strain fields. 
In other words, ${{\mathbf{\dot{F}}}_{\text{I}}}$ can be seen as pure elastic and not influenced by the growth process. 
So, combining  Eqs.\eqref{F} and \eqref{Fdot}, we have 
\begin{equation}
{{\mathbf{\dot{F}}}_{e}}={{\mathbf{\dot{F}}}_{\text{I}}}{{\mathbf{F}}_{e}}
\end{equation}

Next, expanding $\det \left(\mathbf F + \mathbf{\dot{F}}_{\text{I}}\right)$ as follows 
\begin{equation}
\det \left( \mathbf{F}+{{{\mathbf{\dot{F}}}}_{\text{I}}} \right)=\det \left( \mathbf{F} \right)\text{+tr}\left( \det \left( \mathbf{F}{{\mathbf{F}}^{\text{-1}}} \right){{{\mathbf{\dot{F}}}}_{\text{I}}} \right) + \ldots,
\end{equation}
we find the incremental incompressibility condition as
\begin{equation}
\text{tr}({{\mathbf{\dot{F}}}_{\text{I}}})=0.
\end{equation}

With a Taylor series expansion, the incremental nominal stress $\mathbf{\dot{S}}$ can be expressed as 
\begin{equation}
\mathbf{\dot{S}}=J\mathbf{F}_{\text{g0}}^{-1}\left(  \bm{\mathcal{A}}_{e} {{{\mathbf{\dot{F}}}}_{e}}-\dot{p}\mathbf{F}_{e}^{-1}\text{+}p\mathbf{F}_{e}^{-1}{{{\mathbf{\dot{F}}}}_{e}}\mathbf{F}_{e}^{-1} \right),
\quad \text{where} \quad \bm{\mathcal{A}}_{e}=\frac{\partial \psi }{\partial {{\mathbf{F}}_{e}}\partial {{\mathbf{F}}_{e}}}
\end{equation}
 is the (fourth-order) referential elasticity  tensor. 
 Because the push-forward form of the incremental nominal stress is ${{\mathbf{\dot{S}}}_{\text{I}}}={{J}^{-1}}\mathbf{F\dot{S}}$, we find
\begin{equation}
\begin{split}
 {{\mathbf{\dot{S}}}_{\text{I}}} = & {{\mathbf{F}}_{e}}\left( \bm{\mathcal{A}}_{e} {{{\mathbf{\dot{F}}}}_{e}} - \dot{p}\mathbf{F}_{e}^{-1}\text{+}p\mathbf{F}_{e}^{-1}{{{\mathbf{\dot{F}}}}_{e}}\mathbf{F}_{e}^{-1} \right) 
\\
  = & 
 \bm{\mathcal{A}}_{e}^{\text{I}} {{\mathbf{\dot{F}}}_{\text{I}}}-\dot{p}\mathbf{I}\text{+}p{{\mathbf{\dot{F}}}_{\text{I}}}, 
\qquad   \text{ where } \quad  \bm{\mathcal{A}}_{e} = {{\mathbf{F}}_{e}}{{\mathbf{F}}_{e}}\frac{\partial \psi }{\partial {{\mathbf{F}}_{e}}\partial {{\mathbf{F}}_{e}}}
\end{split}
\end{equation}
 is the instantaneous elasticity tensor \citep{RN510}. 
 In component form,  
\begin{equation}
{{\dot{S}}_{{Iij}}} = \mathcal A_{{eijkl}}^{I} {{\dot{F}}_{{Ilk}}}-\dot{p}{{\delta }_{ij}} {+} p{{\dot{F}}_{{Iij}}},
\quad 
\mathcal A_{{eijkl}}^{{I}}={{F}_{{ei}\alpha }}{{F}_{{ek}\beta }}\frac{\partial \psi }{\partial {{F}_{{ej}\alpha }}\partial {{F}_{{el}\beta }}},
\end{equation}
The non-zero components of $ \bm{\mathcal{A}}_{e}^{\text{I}}$ in the coordinate system aligned with the principal axes are  \citep{RN510} 
\begin{align}
& \mathcal A_{eiijj}^{{I}}= \mathcal A_{ejjii}^{{I}}={{\lambda }_{i}}{{\lambda }_{j}}{{\psi }_{ij}}, 
 && \notag \\[6pt]
& \mathcal A_{eijij}^{{I}}=\frac{{{\lambda }_{i}}{{\psi }_{i}}-{{\lambda }_{j}}{{\psi }_{j}}}{\lambda _{i}^{2}-\lambda _{j}^{2}}\lambda _{i}^{2}, 
 && {{\lambda }_{i}}\ne {{\lambda }_{j}}, \notag \\[6pt]
& \mathcal A_{eijji}^{{I}}= \mathcal A_{{ejiij}}^{{I}}= \mathcal A_{eijij}^{{I}}-{{\lambda }_{i}}{{\psi }_{i}}, &&  i\ne j, 
\label{Ae}
\end{align}
where $\psi_i = \partial \psi / \partial \lambda_i$, $\psi_{ij} = \partial^2 \psi / \partial \lambda_i \partial \lambda_j$. 

The equations of incremental equilibrium are 
\begin{equation} \label{divS}
\text{Div }\mathbf{\dot{S}} = \mathbf 0, \qquad \text{or} \quad  \text{div }{{\mathbf{\dot{S}}}_\text{I}} = \mathbf 0.
\end{equation}

Finally, the increment nominal stress and the displacement satisfy the boundary conditions
\begin{equation}
\mathbf{\dot{S}}_\text{I}^\text{T}\mathbf{n} = \mathbf 0,\qquad {{\left( \text{d} \mathbf x \right)}^{\bm{\cdot} }} = \mathbf 0.
\end{equation}


\subsection{Incremental field in the tubular organ}


We write the incremental displacement field as
\begin{equation}
\mathbf{\dot{x}}=u\left( r,\theta  \right){{\mathbf{e}}_{r}}+v\left( r,\theta  \right){{\mathbf{e}}_{\theta}},
\end{equation}
resulting in the following incremental displacement gradient tensor
\begin{equation}
{{\mathbf{\dot{F}}}_{\text{I}}}=\frac{\partial u}{\partial r}{{\mathbf{e}}_{r}}\otimes{{\mathbf{e}}_{r}}+\frac{1}{r}\left( \frac{\partial u}{\partial \theta }-v \right){{\mathbf{e}}_{r}}\otimes{{\mathbf{e}}_{\theta}}+\frac{\partial v}{\partial r}{{\mathbf{e}}_{\theta}}\otimes{{\mathbf{e}}_{r}}+\frac{1}{r}\left( \frac{\partial v}{\partial \theta }-u \right){{\mathbf{e}}_{\theta}}\otimes{{\mathbf{e}}_{\theta}}.
\end{equation}
Then the incremental equilibrium equations \eqref{divS}$_2$ read 
\begin{equation}
\frac{\partial {{{\dot{S}}}_{{I}rr}}}{\partial r}+\frac{1}{r}\frac{\partial {{{\dot{S}}}_{{I}\theta r}}}{\partial \theta }+\frac{{{{\dot{S}}}_{{I}rr}}-{{{\dot{S}}}_{{I}\theta\theta}}}{r}=0, \qquad
\frac{\partial {{{\dot{S}}}_{{I}r \theta}}}{\partial r}+\frac{1}{r}\frac{\partial {{{\dot{S}}}_{{I}\theta\theta}}}{\partial \theta }+\frac{{{{\dot{S}}}_{{I}r\theta}}+{{{\dot{S}}}_{{I}\theta r}}}{r}=0. \end{equation}

We now seek a solution  in the form
\begin{equation}
\begin{split}
& \left\{ u,v,\dot{p} \right\}=\left\{ U\left( r \right),V\left( r \right), \text i n P\left( r \right) \right\}{{\text{e}}^{\text{i}n\theta}}, \\ 
& \left\{ {{{\dot{S}}}_{{I}rr}},{{{\dot{S}}}_{\text{I}r\theta}},{{{\dot{S}}}_{{I}\theta r}},{{{\dot{S}}}_{{I}\theta\theta}} \right\}=\left\{ {{\Sigma }_{rr}}\left( r \right),{{\Sigma }_{r\theta}}\left( r \right),{{\Sigma }_{\theta r}}\left( r \right),{{\Sigma }_{\theta\theta}}\left( r \right) \right\}{{\text{e}}^{\text{i} n \theta}}, \\ 
\end{split}
\end{equation}
where $n$ is the wrinkle number in the circumferential direction, and $U$, $V$, $P$, ${{\Sigma }_{ij}}$ are functions of $r$ only. 
This mechanical field describes a sinusoidal pattern along the circumferential direction, with amplitude variations along the radial direction. 
Next we introduce the incremental displacement-traction vector $\bm{\eta }$ as
\begin{equation}
\bm{\eta }={{\left[ U,V,\text{i}r{{\Sigma }_{rr}},\text{i}r{{\Sigma }_{r\theta}} \right]}^{\text{T}}},
\end{equation}
and find that the governing equations can be put in the \emph{Stroh form} as
\begin{equation}
\frac{\text{d}}{\text{d}r}\bm{\eta }\left( r \right)=\frac{\text{i}}{r} \begin{bmatrix}
{{\mathbf{G}}_{1}}\left( r \right) & {{\mathbf{G}}_{2}}\left( r \right)  \\[4pt]
{{\mathbf{G}}_{3}}\left( r \right) & \mathbf{\bar{G}}_{1}^{\text{T}}\left( r \right)  \\
\end{bmatrix} \bm{\eta }\left( r \right),
\label{stroh-eqn}
\end{equation}
where the components of the $2\times 2$ sub-blocks are
\begin{equation}
\begin{split}
 & {{\mathbf{G}}_{1}}\left( r \right)= \begin{bmatrix}
\text{i} & -n  \\
-n\left( 1-\sigma  \right) & -\text{i}\left( 1-\sigma  \right)  \\
\end{bmatrix},
\qquad 
{{\mathbf{G}}_{2}}\left( r \right)=  \begin{bmatrix}
\text{0} & 0  \\
0 & {-1}/{\omega }  \\
\end{bmatrix},  \\[8pt] 
& {{\mathbf{G}}_{3}}\left( r \right)= \begin{bmatrix}
\begin{bmatrix}
2\beta +2\omega \left( 1-\sigma  \right)\\+{{n}^{2}}\left[ \gamma -\alpha {{\left( 1-\sigma  \right)}^{2}} \right]
\end{bmatrix}
 & n\left[ 2\beta +\gamma +\omega \left( 1-{{\sigma }^{2}} \right) \right]  \\
-\text{i}n\left[ 2\beta +\gamma +\omega \left( 1-{{\sigma }^{2}} \right) \right] & 
\begin{bmatrix}
\gamma -\omega {{\left( 1-\sigma  \right)}^{\text{2}}}\\\text{+2}{{n}^{2}}\left[ \beta +\omega \left( 1-\sigma  \right) \right]  
\end{bmatrix}\\
\end{bmatrix} ,  \\ 
\end{split}
\end{equation}
with, in general,
\begin{equation}
\begin{split}
&\omega = \mathcal{A}_{e\theta r \theta r}^{\text{I}},
\qquad 
\sigma =\frac{{{\sigma }_{rr}}}{\omega },
\qquad 
\gamma =\mathcal{A}_{{er}\theta\theta r}^{\text{I}},\\[6pt] 
&2\beta={\mathcal{A}_{{errrr}}^{\text{I}}+\mathcal{A}_{\theta\theta\theta\theta}^{{I}}-2\mathcal{A}_{{err}\theta\theta}^{{I}}-2\mathcal{A}_{{er}\theta\theta r}^{{I}}}.\\
\end{split}
\end{equation}
According to Eq.\eqref{Ae}, we have for the initially stressed growing neo-Hookean model \eqref{neo}, 
\begin{equation}
\omega =\mu \lambda _{e}^{-2},\qquad \gamma =\mu \lambda _{e}^{2},\qquad \beta =\frac{\mu }{\text{2}}\left( \lambda _{e}^{2}+\lambda _{e}^{-2} \right),
\end{equation}
where 
\begin{equation}
\lambda_e = \sqrt{\frac{ {{\tau }_{\mit{\Theta \Theta}}}+{{p}_{\bm{\tau}}} }{\mu }}\: \frac{r}{{g_\theta}R}
\end{equation} 
is the circumferential stretch ratio of the elastic deformation gradient tensor ${{\mathbf{F}}_{{e}}}=\mathbf{FF}_{0}^{-1}\mathbf{F}_{{g}}^{-1}$. 


\subsection{The surface impedance method}
\label{section5.3}


The surface impedance method was first proposed by  \cite{ RN598} to investigate wave propagation in inhomogeneous solids and later generalized to study of the stability of inhomogeneously deformed solids \citep{RN384, RN528, RN527}. 

The main result is that the critical growth-induced instability state is reached once the\textit{ inner surface impedance matrix} ${{\mathbf{Z}}^{i}}\left( {{r}} \right)$ satisfies the equation 
\begin{equation}
\det \left( {{\mathbf{Z}}^{i}}\left( {{r}_{o}} \right) \right)={0},
\label{detZ}
\end{equation}
where ${{\mathbf{Z}}^{i}}\left( {{r}} \right)$ is obtained by using the boundary condition ${{\left[ {{\Sigma }_{rr}}\left( r_i \right),{{\Sigma }_{r\theta}}\left( r_i \right) \right]}^{\text{T}}}=\textbf{0}$ on the inner surface. To find ${{\mathbf{Z}}^{i}}\left( {{r}_{o}} \right)$ we must integrate numerically the following Riccati differential equation for $\mathbf Z^i$,
\begin{equation}
\frac{\text{d}{{\mathbf{Z}}^{i}}}{\text{d}r}=\frac{1}{r}\left[ {{\mathbf{Z}}^{i}}{{\mathbf{G}}_{2}}{{\mathbf{Z}}^{i}}+{{\mathbf{G}}_{3}}+\text{i}\mathbf{\bar{G}}_{1}^{\text{T}}{{\mathbf{Z}}^{i}}+\text{i}{{\mathbf{Z}}^{i}}{{\mathbf{G}}_{1}} \right],
\end{equation}
through the thickness, from $r=r_i$ with the initial boundary condition ${{\mathbf{Z}}^{i}}\left( {{r}_{i}} \right)= \mathbf 0$, to $r=r_o$ with Eq.\eqref{detZ} as the target.

Once the target is reached, we obtain the corresponding critical value of differential growth ratio ${{{g}_{\theta }}}/{{g_r}}$ and ${{\mathbf{Z}}^{i}}\left( {{r}_{o}} \right)$, and also the shape of the outer surface from the following ratio,
\begin{equation}
\frac{U\left( {{r}_{o}} \right)}{V\left( {{r}_{o}} \right)}=-\frac{Z_{12}^{i}\left( {{r}_{o}} \right)}{Z_{11}^{i}\left( {{r}_{o}} \right)}=-\frac{Z_{22}^{i}\left( {{r}_{o}} \right)}{Z_{21}^{i}\left( {{r}_{o}} \right)}.
\end{equation}

Finally, to determine the  through-thickness incremental displacement field of the tube, we solve simultaneously the following equations for $\mathbf{U}\left( r \right)={{\left[ {U}\left( r \right),{V}\left( r \right) \right]}^{\text{T}}}$ and the \textit{outer conditional impedance matrix} ${{\mathbf{Z}}^{o}}$, see \cite{RN527},
\begin{equation}
\frac{\text{d}\left( {{\mathbf{Z}}^{o}}\mathbf{U} \right)}{\text{d}r}=\frac{1}{r}{{\mathbf{G}}_{3}}\mathbf{U}+\frac{\text{i}}{r}\mathbf{\bar{G}}_{1}^{\text{T}}{{\mathbf{Z}}^{o}}\mathbf{U}, \quad
 \frac{\text{d}{{\mathbf{Z}}^{o}}}{\text{d}r}=\frac{1}{r}\left[ {{\mathbf{Z}}^{o}}{{\mathbf{G}}_{2}}{{\mathbf{Z}}^{o}}+{{\mathbf{G}}_{3}}+\text{i}\mathbf{\bar{G}}_{1}^{\text{T}}{{\mathbf{Z}}^{o}}+\text{i}{{\mathbf{Z}}^{o}}{{\mathbf{G}}_{1}} \right],
\end{equation}
with the initial boundary conditions: ${U\left( {{r}_{o}} \right)}/{V\left( {{r}_{o}} \right)}=-{Z_{12}^{i}\left( {{r}_{o}} \right)}/{Z_{11}^{i}\left( {{r}_{o}} \right)}$ and ${{\mathbf{Z}}^{o}}\left( {{r}_{o}} \right)=\mathbf{0}$.

\subsection{Results}



\subsubsection{Specialisation to a non-growing, residually stressed tube}


First we check that we recover the  instability analysis of \citep{RN384} when ${{\mathbf{F}}_{g}}=\mathbf{I}$. 
In that case, the tube is not growing and the instability is triggered by increasing the amplitude of the initial stress in Eq.\eqref{f} until the critical initial stress amplitude ${{\alpha }_{\text{cr}}}$ is reached. 

\begin{figure}
	\centering	
	\includegraphics[width=0.49\textwidth]{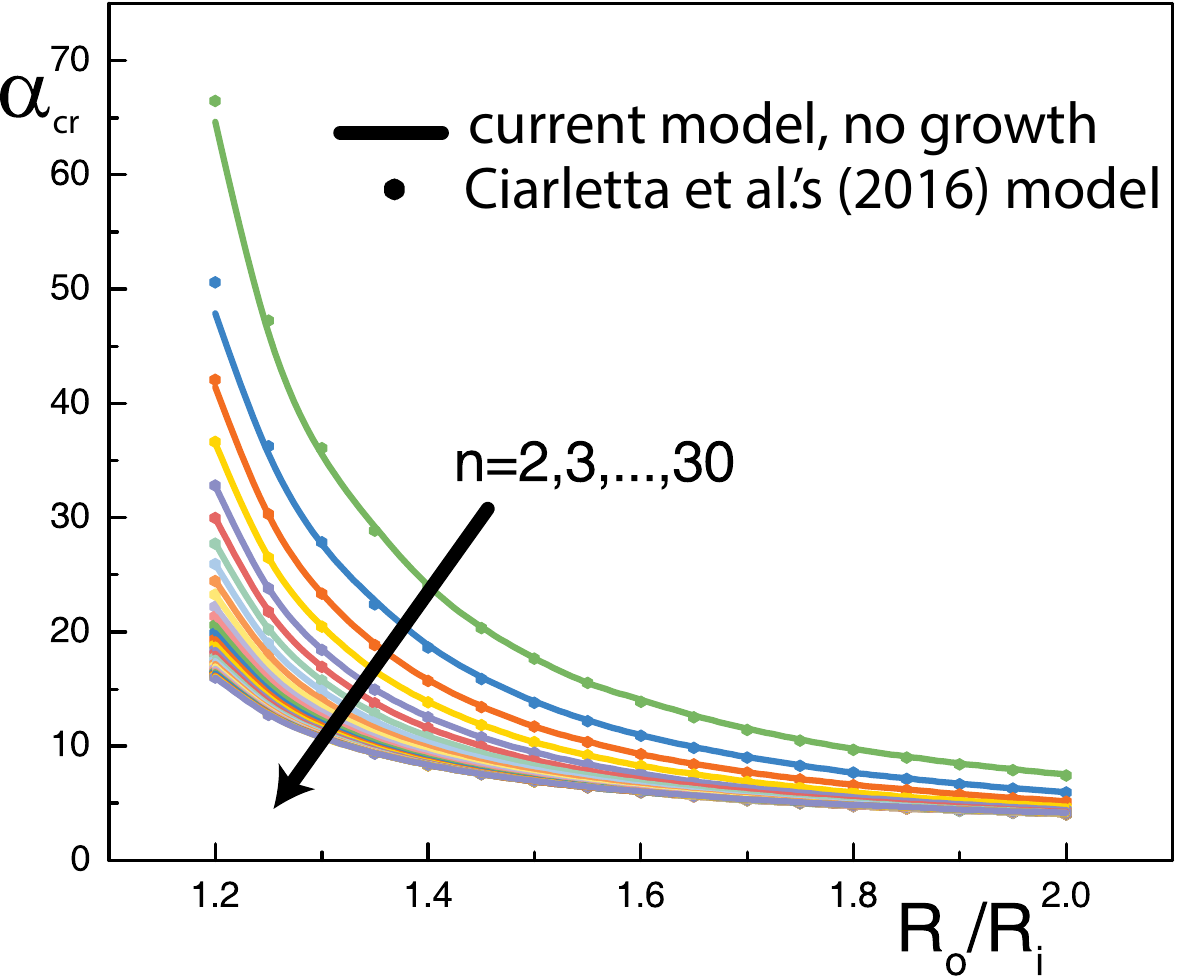}
	\includegraphics[width=0.49\textwidth]{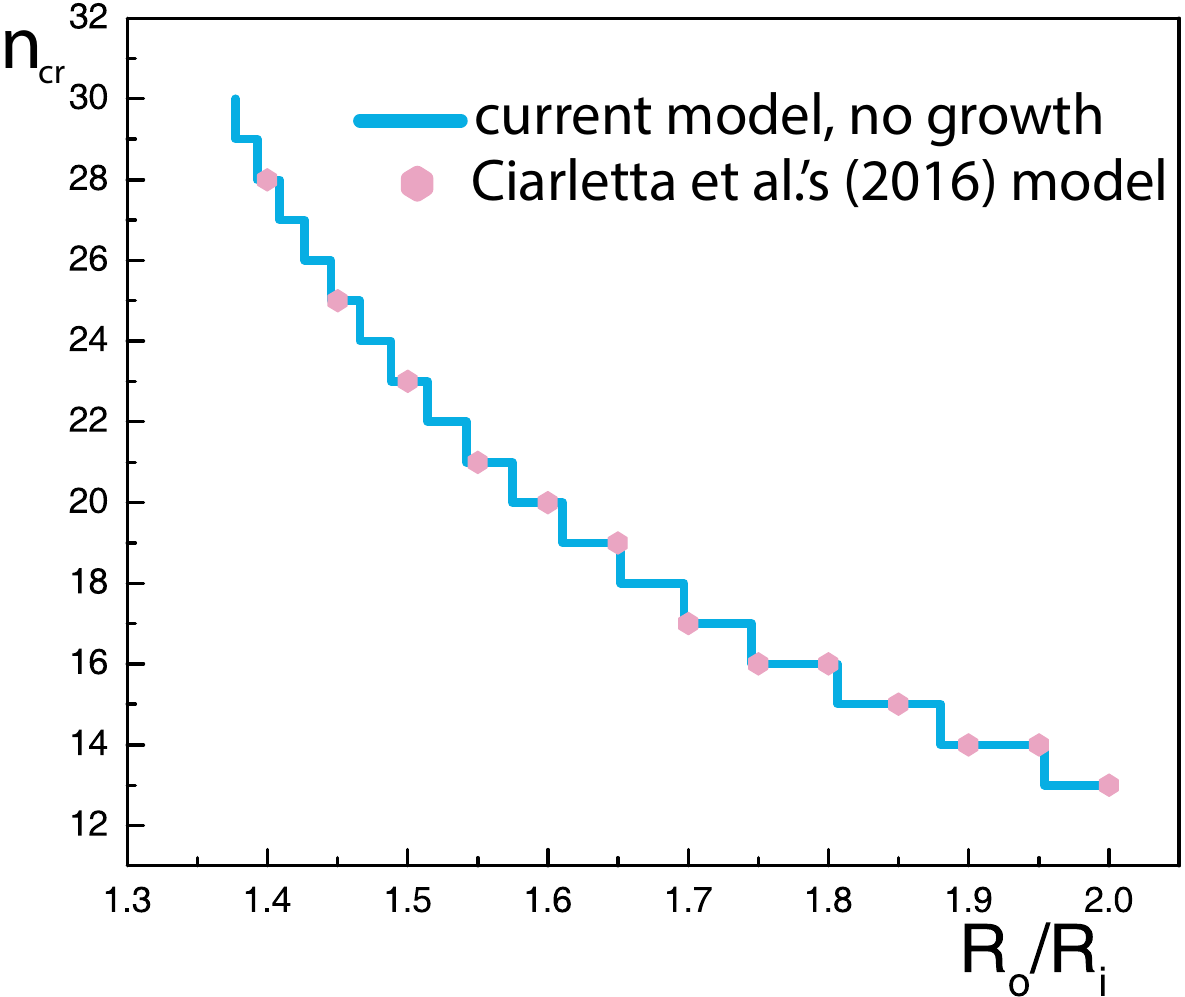}
	\caption{Critical initial stress magnitude and critical wave number against the thickness ratio, in the case of a non-growing residually stressed tube.}
	\label{Fig:10}
\end{figure}

Figure \ref{Fig:10} shows the same results as those by  calculated by \citep{RN384}, which validates our code for the MMD growth model when ${{\mathbf{F}}_{g}}=\mathbf{I}$. 
Figure  \ref{Fig:10}(a) shows the lines of critical magnitude ${{\alpha }_{\text{cr}}}$ of the residual stress against the tube aspect ratio $R_o/R_i$, for different wrinkle numbers $n$. 
For a given ${{R_o}}/{{R_i}}$, there exists a minimal critical stress magnitude, which we record; then by varying ${{R_o}}/{{R_i}}$, we construct the bottom envelope line. 
We also record the corresponding critical wrinkle number, to create Figure \ref{Fig:10}(b), showing the variations of $n_\text{cr}$ with $R_o/R_i$. 
There we see that for a tube with a larger wall thickness, the wrinkle number is smaller, consistent with the results of \cite{RN384}.

These results show that the instability of a soft tissue with residual stress is directly related to its geometry and to the magnitude of the residual stress. 
The instability analysis demonstrates that when wrinkles are present, the magnitude $\alpha_\text{cr}$ of the residual stress can be found \emph{non-destructively} by observing its shape and counting the number of wrinkles. 
There is no need to rely on the multiple decomposition method and on cutting the tube.

Now we introduce differential growth to see how initial stress, geometry, and growth affect pattern formation. 


\subsubsection{Instability analysis for growing initially stressed tube}


Here we adopt the constant growth rate model presented by \cite{RN570} and the logarithmic distribution of initial stress in the reference configuration given by Eq.\eqref{f}. 
Hence, the components of the growth deformation gradient tensor at time $t$ are 
\begin{equation}
g_r\left( t \right)=1+ \dot g_r t, \qquad
g_{\theta}\left( t \right)=1+ \dot g_\theta t, 
\end{equation}
where ${{\dot{g}}_{r}},\ {{\dot{g}}_{\theta }}$ are the constant growth rates in the radial and circumferential directions, respectively. 
We call  $v={{{{\dot{g}}}_{\theta }}}/{{{{\dot{g}}}_{r}}}$ the upper limit of the differential growth ratio ${{g_\theta}}/{{g_r}}$ with time. 


Here we take $\alpha$, the non-dimensional amplitude measure of the initial stress $\bm{\tau}$ and ${{R_o}}/{{R_i}}$, the initial relative wall thickness, as conditional parameters, and take the differential growth ratio ${{g_\theta}}/{{g_r}}$ as critical parameter.

We first analyse the case of varying $\alpha $ and fixed initial relative wall thickness, ${{R_o}}/{{R_i}}=2.0$. 
We found in the previous section that ${{\alpha }_{\text{cr}}}=4.02$ for ${{R_o}}/{{R_i}}=2.0$ when there is no growth. 
Here we look in turn at the cases when $\alpha =0$ (no initial residual stress), $1.0, 2.0, 4.0$, and when buckling occurs on the inner surface of the tube due to relative radial growth (RRG), when ${{g_\theta}}/{{g_r}}<1$. 
To fix the ideas we take $v=0.1$ and find numerically the critical differential growth extent ${{g_\theta}}/{{g_r}}$ such that Eq.\eqref{detZ} is satisfied.

\begin{figure}
	\centering	
	\includegraphics[width=0.9\textwidth]{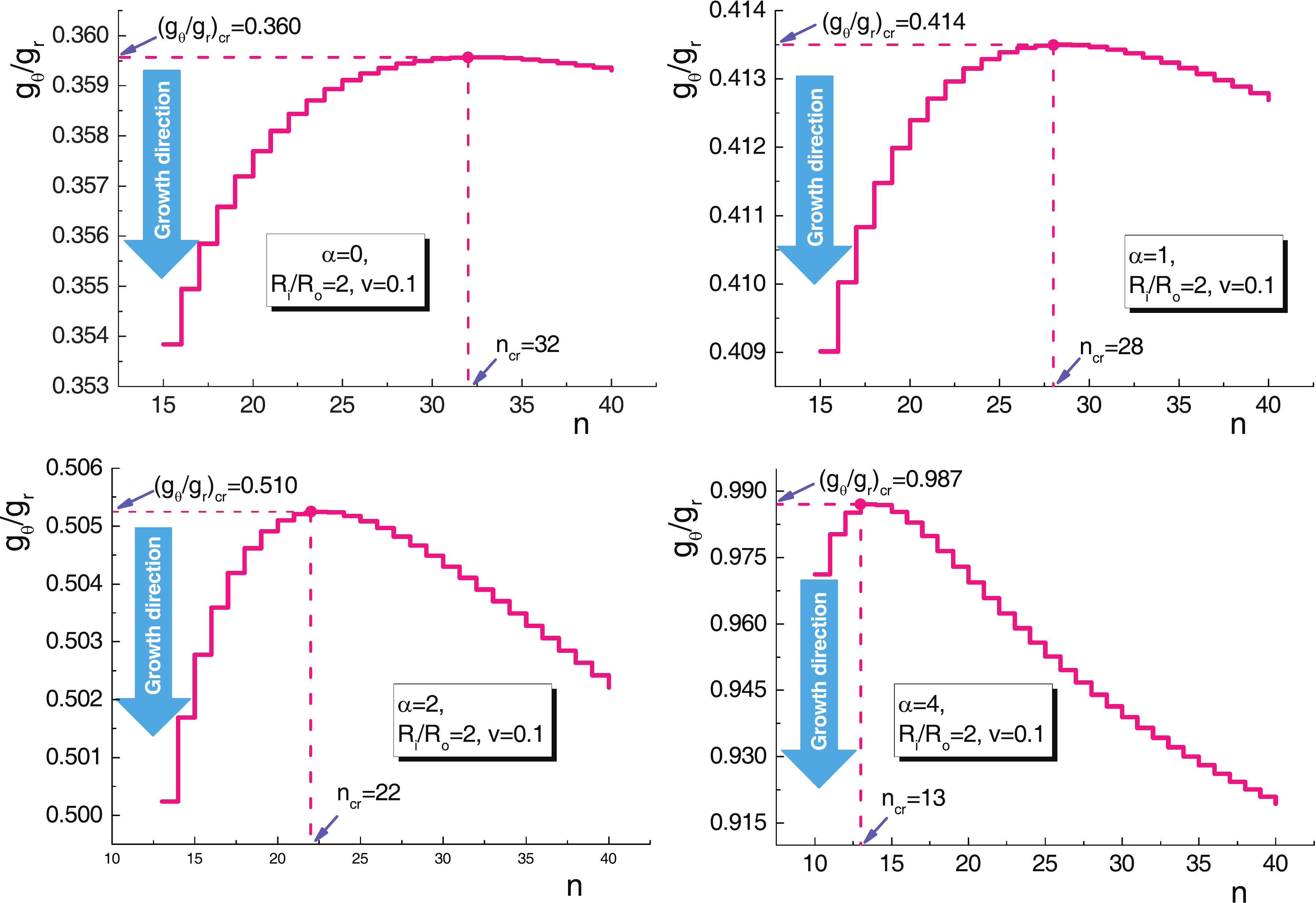}
	\caption{Relationship between the wrinkle number $n$ and the differential growth ratio ${{\left( {{g_\theta}}/{{g_r}} \right)}_{\text{cr}}}$ for different initial residual stress levels $\alpha = 0,1,2,3,4$, when $R_o/R_i=2$. 
	\color{black}	The peak point of the $n \sim g_\theta/g_r$  curve indicates the onset of wrinkles at a critical value of  $g_\theta/g_r$ with a critical half-wave number $n_\text{cr}$.
	\color{black}
	}
	\label{Fig:11}
\end{figure}

Figure \ref{Fig:11} reveals that there exists a different maximal differential growth ratio for each  initial configuration and that the level of initial stress has a significant influence on the instability pattern. 
Hence when there is no initial residual stress ($\alpha=0$), we find ${{\left( {{g_\theta}}/{{g_r}} \right)}_{\text{cr}}}=0.360$, which indicates a large difference in the growth rates (and then there are $n_\text{cr} = 32$ wrinkles), while when $\alpha=4.0$, the growth rates are almost equal: ${{\left( {{g_\theta}}/{{g_r}} \right)}_{\text{cr}}}=0.987$ ($n_\text{cr} = 13$ wrinkles) indicating that only a small relative radial growth process is required to accumulate more residual stress and induce instability. 
Hence relative radial growth makes it easier to induce instability. 
This is particularly true when the initial stress state is $\alpha =4$, very close to the ${{\alpha }_{\text{cr}}}=4.02$ found for non-growing tubes. 
This state is very unstable as a small relative radial growth can induce instability.

\begin{figure}
	\centering	
	\includegraphics[width=0.9\textwidth]{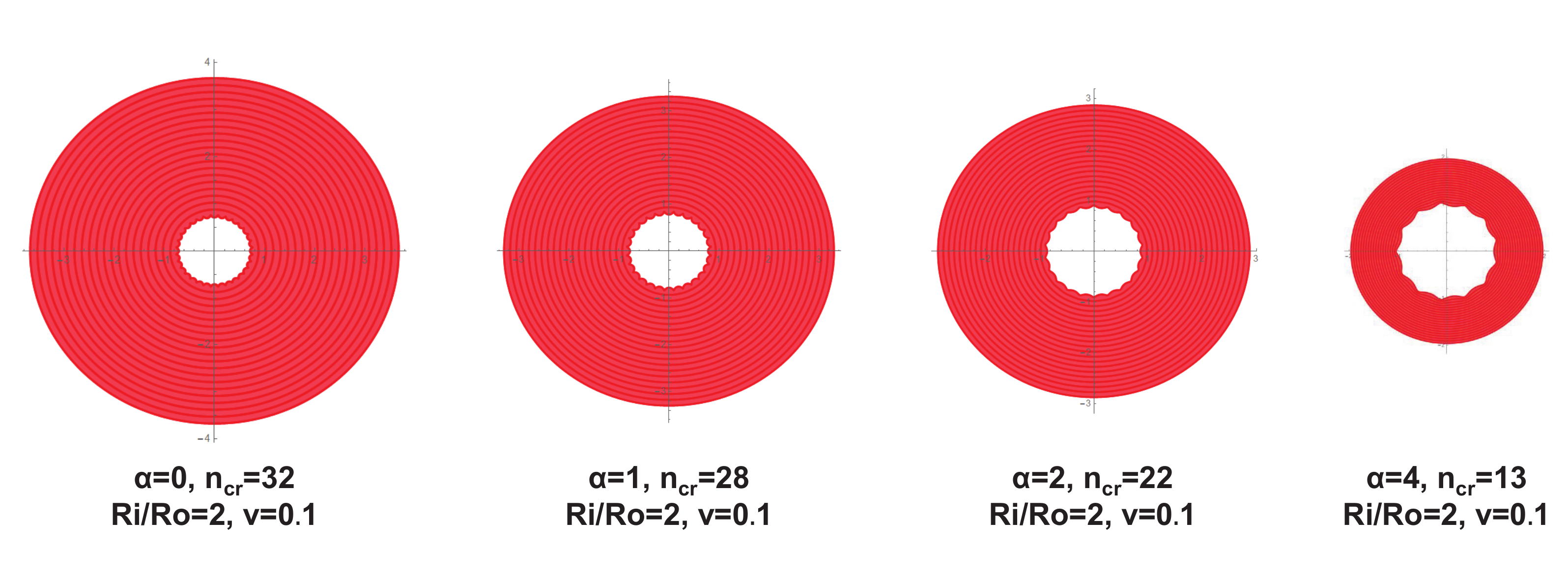}
	\caption{Morphology of unstable states from different initially stressed states  for different initial residual stress levels $\alpha = 0,1,2,3,4$, when the initial geometry of the tube is given by $R_o/R_i=2$.}
	\label{Fig:12}
\end{figure}

Once we know the critical wave number ${{n}_{\text{cr}}}$ and the critical differential growth extent ${{\left( {{g_\theta}}/{{g_r}} \right)}_{\text{cr}}}$, we can compute the whole incremental mechanical displacement field as explained in Section \ref{section5.3}. 
Figure \ref{Fig:12} presents the resulting unstable morphologies for the same levels of initial residual stress as in Figure \ref{Fig:11}.

\begin{figure}
	\centering	
	\includegraphics[width=0.9\textwidth]{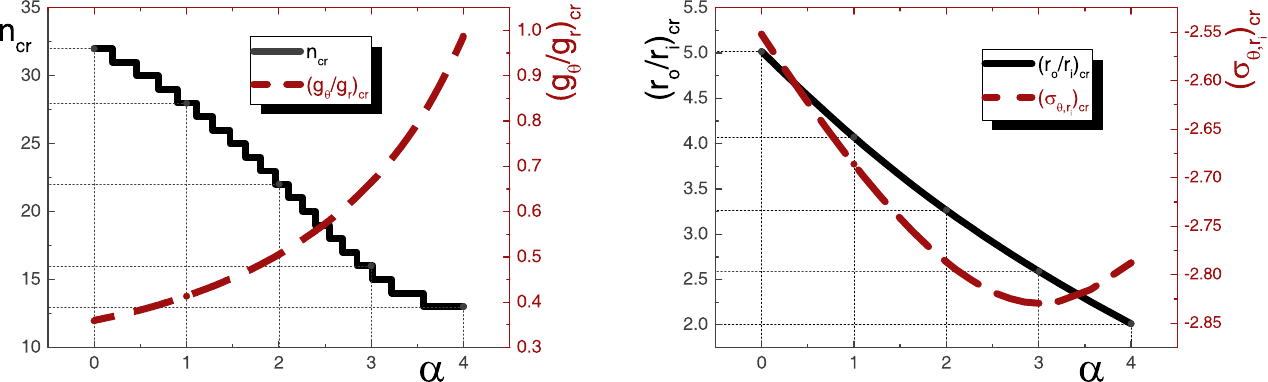}
	\caption{(a) Relationship between initial stress level $\alpha$ and  critical wrinkle number $n_\text{cr}$ or the critical differential growth ratio ${{\left( {{g_\theta}}/{{g_r}} \right)}_{\text{cr}}}$  for the growth-induced unstable state, for $R_o/R_i=2$. (b) The relationship between the initial stress level and the critical geometric size or the critical circumferential residual stress on the inner surface for the growth-induced instable state.}
	\label{Fig:13}
\end{figure}

For the same initial geometric size ${{R_o}}/{{R_i}}=2$, we calculate the critical differential growth extent ${{\left( {{g_\theta}}/{{g_r}} \right)}_{\text{cr}}}$ when the initial stress level varies continuously between 0 and 4. 
The resulting Figure \ref{Fig:13}(a) shows the effect of the initial stress level on growth-induced instability. 
For an initially stressed material, the conditional parameters $\alpha $ and ${{\left( {{g_\theta}}/{{g_r}} \right)}_{\text{cr}}}$ are positively correlated. Additionally, because the time limit of the differential growth ratio ${{\left( {{g_\theta}}/{{g_r}} \right)}_{\text{cr}}}$ is equal to the relative differential growth rate $v$, the red line in Figure \ref{Fig:13}(a) can also be seen as giving the critical (or minimal) relative differential growth rate ${{v}_{\text{cr}}}$. 
In other words, if the ratio of the two constant growth rates is less than  ${{v}_{\text{cr}}}$, then no instability will occur in the tissue.

Figure \ref{Fig:13}(b) shows the influence of $\alpha$ on the ratio of the outer radius to the inner radius ${{{r}_{o}}}/{{{r}_{i}}}$ in the current configuration and on the circumferential residual stress found on the inner surface $\sigma_{\theta}(r_i)$. 

\begin{figure}
	\centering	
	\includegraphics[width=0.46\textwidth]{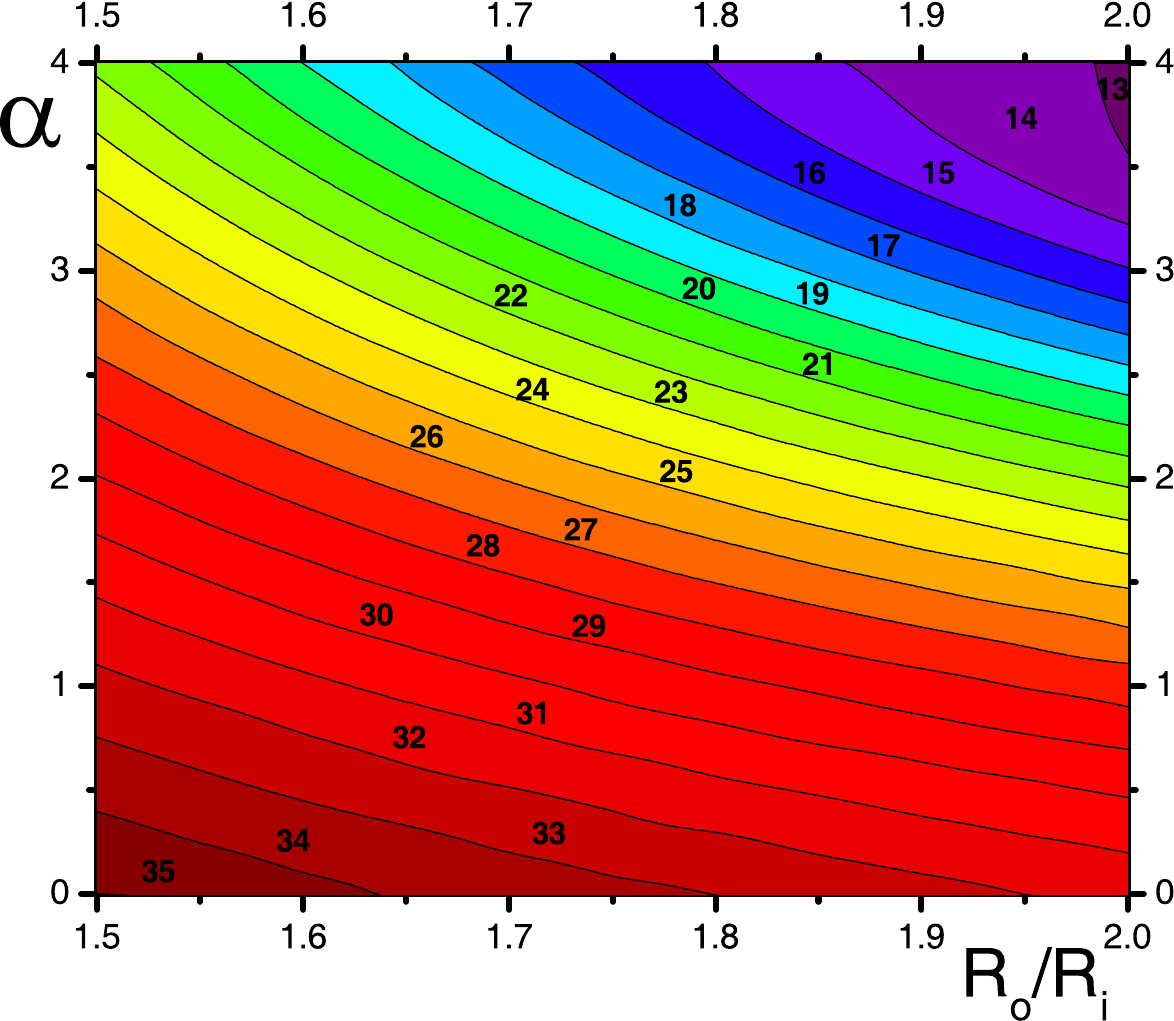} \quad
	\includegraphics[width=0.5\textwidth]{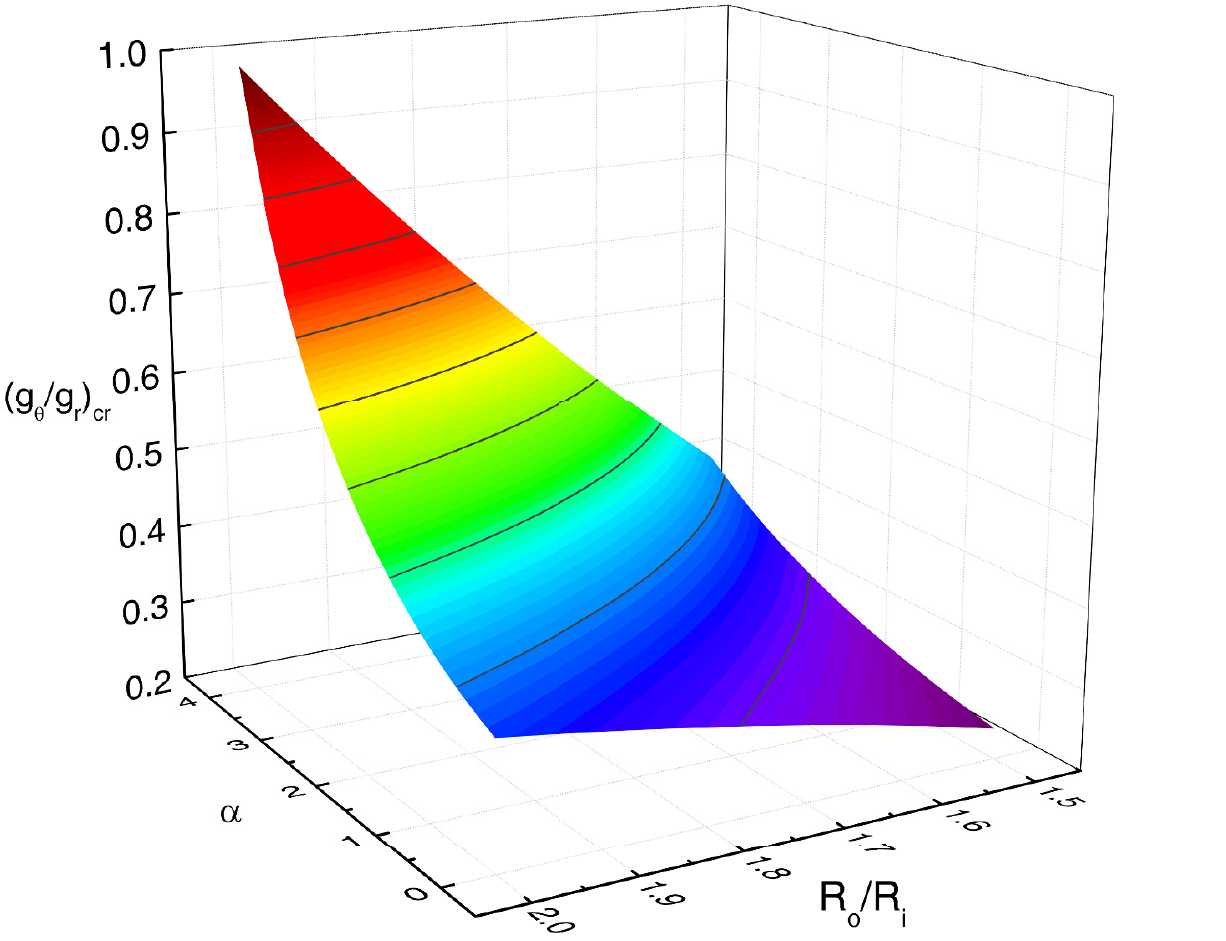}
	\caption{(a)  Critical wave number ${{n}_{\text{cr}}}$ and (b)  critical differential growth extent ${{\left( {{g_\theta}}/{{g_r}} \right)}_{\text{cr}}}$for initially stressed materials}
	\label{Fig:14}
\end{figure}

Considering now the influence of the other conditional parameter, the relative wall thickness ${{R_o}}/{{R_i}}$, we conduct the instability analysis when it is varying continuously between 1.5 and 2.0.
Figure \ref{Fig:14}(a) shows the values of the resulting critical wrinkle number ${{n}_{\text{cr}}}$ and of the critical differential growth extent ${{\left( {{g_\theta}}/{{g_r}} \right)}_{\text{cr}}}$ when the magnitude of the  initially residual stress varies between 0 and 4. 
We see that a higher initial stress amplitude $\alpha$  and a thicker initial relative wall thickness ${{R_o}}/{{R_i}}$ lead to less wrinkles and a lower differential growth rate ratio. 

We conclude that for an arbitrary initially stressed state, the initial conditional parameters such as the initial non-dimensional amplitude measure $\alpha $ of the initial stress $\bm{\tau}$ and the initial relative wall thickness ${{R_o}}/{{R_i}}$ are vital to determining the onset of critical instability patterns. 
\begin{figure}
	\centering	
	\includegraphics[width=0.9\textwidth]{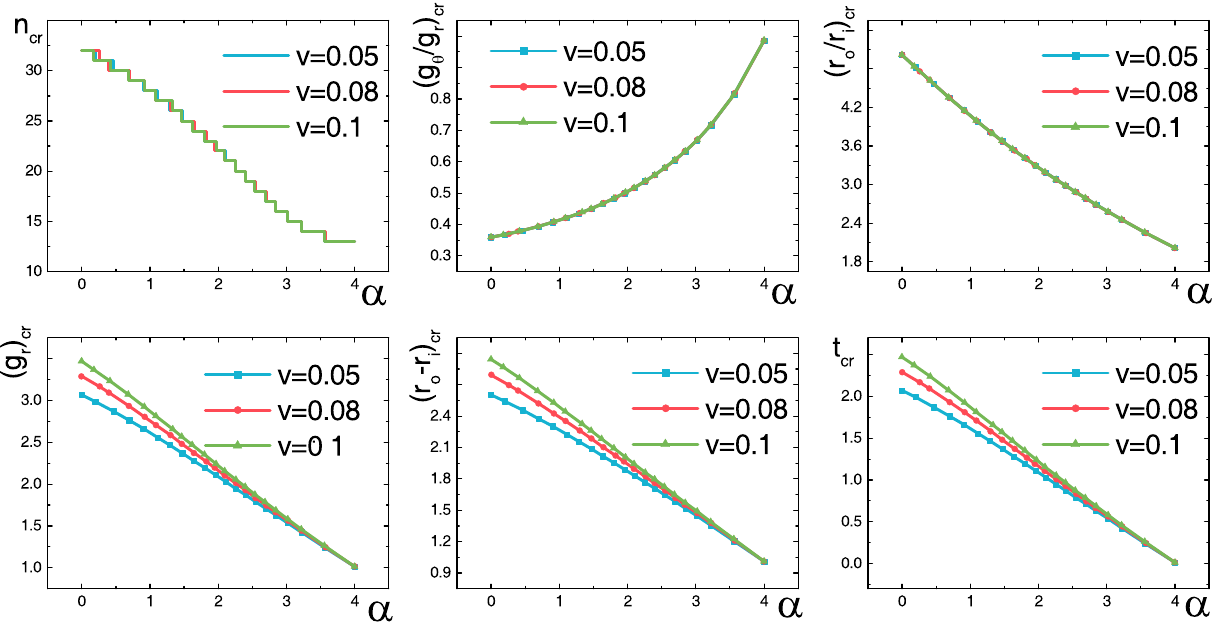}
	\caption{Influence of the differential growth rate $v$ on instability patterns, when the initial relative wall thickness is ${{R_o}}/{{R_i}}=2$ and the time-scale is equal to 1.}
	\label{Fig:15}
\end{figure}

We find that the constant different growth rate ratio $v$ is also a decisive parameter to evaluate whether instability occurs. 
In Figure \ref{Fig:15} we display the results of our calculations for different growth rate ratios $v=0.05$, $0.08$, $0.1$ when initial stress level $\alpha$ varies from 0 to 4 and when the initial relative wall thickness is fixed as ${{R_o}}/{{R_i}}=2$.
Here, we find that  the critical wave number ${{n}_{\text{cr}}}$, the critical differential growth extent ${{\left( {{g_\theta}}/{{g_r}} \right)}_{\text{cr}}}$ and the critical relative wall thickness ${{\left( {{{r}_{o}}}/{{{r}_{i}}} \right)}_{\text{cr}}}$ of the growing initially stressed tube do not depend on the differential growth rate ratio $v$, as the three curves superpose. 
However, the growth factor ${{\left( {g_r} \right)}_{\text{cr}}}$, the absolute wall thickness ${{\left( {{r}_{o}}-{{r}_{i}} \right)}_{\text{cr}}}$ and the growth time required for instability ${{t}_{\text{cr}}}$ do depend on $v$.


\section{Discussion and conclusion}


In this  work, we  proposed a modified multiplicative decomposition (MMD) growth model to simulate  growth  in a general way, starting with an initially stressed reference configuration instead of the inaccessible stress-free state used in  the MD model \citep{ RN101}. 

In Section \ref{section2}, we presented the kinematic description for initially stressed growing matter.
We introduced a virtual stress-free  (VSF) configuration corresponding to the reference configuration to show the unconstrained or incompatible discrete state and we assumed that the constitutive equation for the stress-free state is known and invertible. 
Based on the current growth model, we derived the free energy function for the material.

Based on the neo-Hookean model, we established a constitutive equation for initially stressed growing matter in Section \ref{section3}. 
By assuming  axisymmetric growth deformation and elastic incompressibility, we \color{black} obtained \color{black} the releasing initial stress deformation and further, the Cauchy stress in the current configuration. 

In Section \ref{section4}, we treated the example of a growing axisymmetric tube with an initial residual stress field $\bm{\tau}$ that is found by using the Airy function method. 
Figure \ref{Fig:6} shows that the MMD growth model recovers the classical volume growth model when there is no initial residual stress ($\bm \tau = \bm 0$). 
Then we showed that residual stress is caused by differential growth. 
Relative radial growth (RRG) accumulates compressive stress on the inner surface, while relative circumferential growth (RCG) generates tensile stress on the inner surface. 
The residual stress induced by incompatible growth can be quantified by the differential growth ratio ${{g_\theta}}/{{g_r}}$. 
In Figures \ref{Fig:7} and \ref{Fig:8}, we calculated the growth-induced residual stress with different initial stress levels and showed the differences between the results of \cite{RN101} (MD growth model) and of our MMD growth model. 
The initial stress clearly has a significant impact on the residual stress distribution in the current configuration. 

Finally in Section \ref{section5} we applied the incremental theory to initially stressed growing matter and derived the incremental field equations for wrinkles in  a growing tube.
With the surface impedance method, we accessed the entire incremental displacement and traction fields. 
In Figure \ref{Fig:10}, we showed that our MMD model recovers the results of \cite{RN384} for instability in a non-growing residually stressed configuration ($\mathbf F_g = \mathbf{I}$). 
We checked that residual stress level and relative wall thickness are two critical factors to determine the critical instability pattern. 
Then we implemented a growth process with a constant growth rate ratio $v$.
We saw in Figure \ref{Fig:14} that the same two  conditional parameters again have a significant impact on the critical instability pattern. 
%
With Figure \ref{Fig:15}, we showed that the differential growth rate ratio $v$ does not affect the critical wave number, the relative wall thickness and the critical differential growth ratio but will generate different absolute wall thickness and critical time requirement. 

These results present a valid approach to tune or control the growth process and the tissue morphology.

In conclusion, this work shows that the multiplicative decomposition (MD) growth model can be modified and enhanced by considering an initial stress in the reference configuration.
\color{black}Compared with the cumulative growth law \citep{goriely2007}, the current MMD growth model proposes a more general framework where the residual stress does not need to stem only from a prior growth process. Even if the residual stress in the initial state stems from growth, the growth process can be analyzed using the current model in one step rather than using Goriely and Ben Amar's multi-step scheme \citep{goriely2007}, with the first step depending on a prescribed unstressed configuration. \color{black}

Hence growth can be expanded to model the growth process between two arbitrary residually stressed states, a situation which is extremely common in real biological growth. 
The modified multiplicative decomposition (MMD) growth model proposed here recovers the MD growth model when there is no initial stress and the residually stressed model when there is no growth. 
Our instability analysis for a cylindrical tube with constant growth rates shows that  wrinkle patterns are related to  initial conditional parameters including the initial stress level and the corresponding geometric size. 

 \color{black} The results demonstrate that the initial residual stress may affect the growth and surface morphology of bio-tissues. We believe that the initial residual stress may also affect the formation of creases, or other post-buckling patterns, symmetry breaking, wrinkle mode transition, period doubling, etc. However, this process cannot be covered by the current incremental analyses but requires non-linear finite element simulations, similar to those conducted by \cite{jin2011creases} (growth without initial stress) or \cite{RN384} (residual stress without growth). This  remains an open topic for future study.\color{black}


\section*{Acknowledgments}

This work was supported by the National Natural Science Foundation of China through grant No.11621062 and No.11772295.

\section*{References}

\bibliographystyle{elsarticle-harv.bst}
\bibliography{MMDG_bib.bib}

\begin{thebibliography}{42}
\expandafter\ifx\csname natexlab\endcsname\relax\def\natexlab#1{#1}\fi
\expandafter\ifx\csname url\endcsname\relax
  \def\url#1{\texttt{#1}}\fi
\expandafter\ifx\csname urlprefix\endcsname\relax\def\urlprefix{URL }\fi

\bibitem[{Balbi et~al.(2015)Balbi, Kuhl, and Ciarletta}]{RN529}
Balbi, V., Kuhl, E., Ciarletta, P., 2015. Morphoelastic control of
  gastro-intestinal organogenesis: Theoretical predictions and numerical
  insights. Journal of the Mechanics and Physics of Solids 78, 493--510.

\bibitem[{Ben~Amar and Goriely(2005)}]{RN90}
Ben~Amar, M., Goriely, A., 2005. Growth and instability in elastic tissues.
  Journal of the Mechanics and Physics of Solids 53~(10), 2284--2319.

\bibitem[{Biryukov(1985)}]{RN598}
Biryukov, S.~V., 1985. Impedance method in the theory of elastic surface waves.
  Soviet Physics Acoustics Ussr 31~(5), 350--354.

\bibitem[{Chen and Eberth(2012)}]{RN488}
Chen, Y.~C., Eberth, J.~F., 2012. Constitutive function, residual stress, and
  state of uniform stress in arteries. Journal of the Mechanics and Physics of
  Solids 60~(6), 1145--1157.

\bibitem[{Chuong and Fung(1983)}]{RN281}
Chuong, C.~J., Fung, Y.~C., 1983. Three-dimensional stress distribution in
  arteries. Journal of Biomechanical Engineering 105~(3), 268--274.

\bibitem[{Chuong and Fung(1986)}]{RN282}
Chuong, C.~J., Fung, Y.~C., 1986. On residual stresses in arteries. Journal of
  Biomechanical Engineering 108~(2), 189--192.

\bibitem[{Ciarletta(2013)}]{RN506}
Ciarletta, P., 2013. Buckling instability in growing tumor spheroids. Physical
  Review Letters 110~(15), 158102.

\bibitem[{Ciarletta et~al.(2016{\natexlab{a}})Ciarletta, Destrade, and
  Gower}]{RN487}
Ciarletta, P., Destrade, M., Gower, A.~L., 2016{\natexlab{a}}. On residual
  stresses and homeostasis: an elastic theory of functional adaptation in
  living matter. Scientific Report 6, 24390.

\bibitem[{Ciarletta et~al.(2016{\natexlab{b}})Ciarletta, Destrade, Gower, and
  Taffetani}]{RN384}
Ciarletta, P., Destrade, M., Gower, A.~L., Taffetani, M., 2016{\natexlab{b}}.
  Morphology of residually stressed tubular tissues: Beyond the elastic
  multiplicative decomposition. Journal of the Mechanics and Physics of Solids
  90, 242--253.

\bibitem[{Cowin(2004)}]{RN260}
Cowin, S.~C., 2004. Tissue growth and remodeling. Annual Review of Biomedical
  Engineering 6, 77--107.

\bibitem[{Cowin(2006)}]{RN478}
Cowin, S.~C., 2006. On the Modeling of Growth and Adaptation. Springer Berlin
  Heidelberg, Berlin, Heidelberg, pp. 29--46.

\bibitem[{Destrade et~al.(2010)Destrade, Murphy, and Ogden}]{RN528}
Destrade, M., Murphy, J.~G., Ogden, R.~W., 2010. On deforming a sector of a
  circular cylindrical tube into an intact tube: Existence, uniqueness, and
  stability. International Journal of Engineering Science 48~(11), 1212--1224.

\bibitem[{Destrade et~al.(2009)Destrade, N\'{i}~Annaidh, and Coman}]{RN527}
Destrade, M., N\'{i}~Annaidh, A., Coman, C.~D., 2009. Bending instabilities of
  soft biological tissues. International Journal of Solids and Structures
  46~(25-26), 4322--4330.

\bibitem[{Du and L\"{u}(2017)}]{RN612}
Du, Y.~K., L\"{u}, C.~F., 2017. Modeling on monitoring the growth and rupture
  assessment of saccular aneurysms. Theoretical and Applied Mechanics Letters
  7~(2), 117--120.

\bibitem[{Eskandari and Kuhl(2015)}]{RN570}
Eskandari, M., Kuhl, E., 2015. Systems biology and mechanics of growth. Wiley
  Interdisciplinary Reviews: Systems Biology and Medicine 7~(6), 401--412.

\bibitem[{Fung(1991)}]{RN523}
Fung, Y.~C., 1991. What are the residual stresses doing in our blood vessels?
  Annals of Biomedical Engineering 19~(3), 237.

\bibitem[{Fung(2013)}]{RN575}
Fung, Y.~C., 2013. Biomechanics: motion, flow, stress, and growth. Springer
  Science \& Business Media.

\bibitem[{Goriely and Amar(2007)}]{goriely2007}
Goriely, A., Amar, M.~B., 2007. On the definition and modeling of incremental,
  cumulative, and continuous growth laws in morphoelasticity. Biomechanics and
  Modeling in Mechanobiology 6~(5), 289--296.

\bibitem[{Gower et~al.(2015)Gower, Ciarletta, and Destrade}]{RN389}
Gower, A.~L., Ciarletta, P., Destrade, M., 2015. Initial stress symmetry and
  its applications in elasticity. Proceedings of the Royal Society A:
  Mathematical, Physical and Engineering Science 471~(2183), 20150448.

\bibitem[{Gower et~al.(2017)Gower, Shearer, and Ciarletta}]{Gower17}
Gower, A.~L., Shearer, T., Ciarletta, P., 2017. A new restriction for initially
  stressed elastic solids. Quarterly Journal of Mechanics and Applied
  Mathematics 70, 455--478.

\bibitem[{Hoger(1997)}]{RN492}
Hoger, A., 1997. Virtual configurations and constitutive equations for
  residually stressed bodies with material symmetry. Journal of Elasticity
  48~(2), 125--144.

\bibitem[{Holzapfel(2000)}]{RN589}
Holzapfel, G., 2000. Nonlinear solid mechanics: a continuum approach for
  engineering. West Sussex, England: John Wiley \& Sons, Ltd.

\bibitem[{Hosford(2010)}]{RN576}
Hosford, W.~F., 2010. Mechanical behavior of materials. Cambridge University
  Press.

\bibitem[{Jin et~al.(2011)Jin, Cai, and Suo}]{jin2011creases}
Jin, L., Cai, S., Suo, Z., 2011. Creases in soft tissues generated by growth.
  EPL (Europhysics Letters) 95~(6), 64002.

\bibitem[{Johnson and Hoger(1993)}]{RN508}
Johnson, B.~E., Hoger, A., 1993. The dependence of the elasticity tensor on
  residual stress. Journal of Elasticity 33~(2), 145--165.

\bibitem[{Johnson and Hoger(1995)}]{RN483}
Johnson, B.~E., Hoger, A., 1995. The use of a virtual configuration in
  formulating constitutive equations for residually stressed elastic materials.
  Journal of Elasticity 41~(3), 177--215.

\bibitem[{Lematre et~al.(2006)Lematre, Feuillard, Le~Clézio, and
  Lethiecq}]{RN588}
Lematre, M., Feuillard, G., Le~Clézio, E., Lethiecq, M., 2006. Modeling of the
  influence of a prestress gradient on guided wave propagation in piezoelectric
  structures. The Journal of the Acoustical Society of America 120~(4),
  1964--1975.

\bibitem[{Li et~al.(2011{\natexlab{a}})Li, Cao, and Feng}]{RN511}
Li, B., Cao, Y.~P., Feng, X.~Q., 2011{\natexlab{a}}. Growth and surface folding
  of esophageal mucosa: a biomechanical model. Journal of Biomechanics 44~(1),
  182--188.

\bibitem[{Li et~al.(2011{\natexlab{b}})Li, Cao, Feng, and Gao}]{RN98}
Li, B., Cao, Y.~P., Feng, X.~Q., Gao, H.~J., 2011{\natexlab{b}}. Surface
  wrinkling of mucosa induced by volumetric growth: Theory, simulation and
  experiment. Journal of the Mechanics and Physics of Solids 59~(4), 758--774.

\bibitem[{L\"{u} and Du(2016)}]{RN374}
L\"{u}, C.~F., Du, Y.~K., 2016. Theoretical modeling for monitoring the growth
  of fusiform abdominal aortic aneurysms using dielectric elastomer capacitive
  sensors. International Journal of Applied Mechanics 8~(7), 1640010.

\bibitem[{MacLaurin et~al.(2012)MacLaurin, Chapman, Jones, and Roose}]{RN505}
MacLaurin, J., Chapman, J., Jones, G.~W., Roose, T., 2012. The buckling of
  capillaries in solid tumours. Proceedings of the Royal Society A:
  Mathematical, Physical and Engineering Sciences 468~(2148), 4123--4145.

\bibitem[{Moulton and Goriely(2011)}]{RN571}
Moulton, D.~E., Goriely, A., 2011. Circumferential buckling instability of a
  growing cylindrical tube. Journal of the Mechanics and Physics of Solids
  59~(3), 525--537.

\bibitem[{Ogden(1984)}]{RN510}
Ogden, R.~W., 1984. Non-linear elastic deformation. Engineering Analysis with
  Boundary Elements 1~(2), 119.

\bibitem[{Rodriguez et~al.(1994)Rodriguez, Hoger, and Mcculloch}]{RN101}
Rodriguez, E.~K., Hoger, A., Mcculloch, A.~D., 1994. Stress-dependent finite
  growth in soft elastic tissues. Journal of Biomechanics 27~(4), 455--467.

\bibitem[{Roux(1894)}]{RN456}
Roux, W., 1894. The problems, methods, and scope of developmental mechanics.
  Biological lectures of the marine biology laboratory, Woods Hole. Ginn,
  Boston, 149--190.

\bibitem[{Schajer(2013)}]{RN572}
Schajer, G.~S., 2013. Practical Residual Stress Measurement Methods. John Wiley
  \& Sons Inc.

\bibitem[{Shams et~al.(2011)Shams, Destrade, and Ogden}]{RN452}
Shams, M., Destrade, M., Ogden, R.~W., 2011. Initial stresses in elastic
  solids: Constitutive laws and acoustoelasticity. Wave Motion 48~(7),
  552--567.

\bibitem[{Skalak et~al.(1996)Skalak, Zargaryan, Jain, Netti, and Hoger}]{RN255}
Skalak, R., Zargaryan, S., Jain, R.~K., Netti, P.~A., Hoger, A., 1996.
  Compatibility and the genesis of residual stress by volumetric growth.
  Journal of Mathematical Biology 34~(8), 889--914.

\bibitem[{Stylianopoulos et~al.(2012)Stylianopoulos, Martin, Chauhan, Jain,
  Diop-Frimpong, Bardeesy, Smith, Ferrone, Hornicek, Boucher, Munn, and
  Jain}]{RN503}
Stylianopoulos, T., Martin, J.~D., Chauhan, V.~P., Jain, S.~R., Diop-Frimpong,
  B., Bardeesy, N., Smith, B.~L., Ferrone, C.~R., Hornicek, F.~J., Boucher, Y.,
  Munn, L.~L., Jain, R.~K., 2012. Causes, consequences, and remedies for
  growth-induced solid stress in murine and human tumors. Proceedings of the
  National Academy of Sciences of the United State of America 109~(38),
  15101--15108.

\bibitem[{Taber(1998)}]{RN451}
Taber, L.~A., 1998. Biomechanical growth laws for muscle tissue. Journal of
  Theoretical Biology 193~(2), 201--213.

\bibitem[{Wang et~al.(2017)Wang, Zhang, and Chen}]{wang2017}
Wang, Y., Zhang, C., Chen, W., 2017. An analytical model to predict material
  gradient and anisotropy in bamboo. Acta Mechanica 228~(8), 2819--2833.

\bibitem[{Wolff(1893)}]{RN595}
Wolff, J., 1893. Das gesetz der transformation der knochen. DMW-Deutsche
  Medizinische Wochenschrift 19~(47), 1222--1224.

\end{thebibliography}







\end{document}